\documentclass[final,1p,times]{elsarticle}

\usepackage{amssymb}
\usepackage{xfrac}
\usepackage{hyperref}       
\usepackage{url}            
\usepackage{booktabs}       
\usepackage{amsfonts,amsmath}       
\usepackage{nicefrac}       
\usepackage{microtype}      
\usepackage{lipsum}
\usepackage{fancyhdr}       
\usepackage{graphicx}       
\usepackage{subcaption}
\usepackage{transparent}
\usepackage{color}

\journal{Journal of Computational Physics}

\begin{document}

\begin{frontmatter}



\title{Using Biot-Savart boundary conditions for unbounded external flow on Eulerian meshes}


\author[inst1]{Gabriel D. Weymouth\corref{cor1}}
\author[inst1]{Marin Lauber\corref{cor2}}
\cortext[cor1]{G.D.Weymouth@tudelft.nl}
\cortext[cor2]{M.Lauber@tudelft.nl}
\affiliation[inst1]{addressline={Faculty of Mechanical Engineering (ME),\\Delft University of Technology}, 
            city={Delft},
            country={Netherlands}}

\begin{abstract}
We introduce a novel boundary condition for incompressible Eulerian simulations formulated using a Biot-Savart vorticity integral that maintains high-accuracy results even when the domain boundary is within a body-length of immersed solid boundaries. The key prerequisite to accurately couple the Biot-Savart condition to the Eulerian velocity and pressure fields is including the influence of the vorticity generated at the immersed boundaries during the incompressible-flow projection step. While the resulting linear operator for the pressure is non-local, it can be efficiently solved by partitioning it into the standard local Poisson operator and the Biot-Savart update. We use oct-tree clustering for the Fast Multi-level Method (FM$\ell M)$ to reduce the computational cost of the evaluation of the Biot-Savart integral on the boundaries of a 3D simulation with $N$ points from $O(N^{5/3})$ to $O(N)$ and show that this has bounded errors. We show that the new method captures the analytical added-mass force of accelerating 2D and 3D plates exactly and matches experimentally measured wake development even when the entire domain only extends 1/2 diameter from the plate. The new method also predicts accurate time-varying forces for a 2D circle and 3D sphere regardless of domain size, while classical boundary conditions require a domain more than 100 times larger in 2D to converge on the new method's result. Finally, we study the highly sensitive 2D deflected wakes produced by high frequency flapping foils to the new boundary conditions and show that truncating the deflected wake within four cord-lengths of the body changes the body force amplitudes by 10-40\%. Doubling the 
wake size recovers the asymptotic results to within 5\%. 
\end{abstract}

\begin{graphicalabstract}
\def\svgwidth{1\columnwidth}
\begingroup%
  \makeatletter%
  \providecommand\color[2][]{%
    \errmessage{(Inkscape) Color is used for the text in Inkscape, but the package 'color.sty' is not loaded}%
    \renewcommand\color[2][]{}%
  }%
  \providecommand\transparent[1]{%
    \errmessage{(Inkscape) Transparency is used (non-zero) for the text in Inkscape, but the package 'transparent.sty' is not loaded}%
    \renewcommand\transparent[1]{}%
  }%
  \providecommand\rotatebox[2]{#2}%
  \newcommand*\fsize{\dimexpr\f@size pt\relax}%
  \newcommand*\lineheight[1]{\fontsize{\fsize}{#1\fsize}\selectfont}%
  \ifx\svgwidth\undefined%
    \setlength{\unitlength}{508.29994502bp}%
    \ifx\svgscale\undefined%
      \relax%
    \else%
      \setlength{\unitlength}{\unitlength * \real{\svgscale}}%
    \fi%
  \else%
    \setlength{\unitlength}{\svgwidth}%
  \fi%
  \global\let\svgwidth\undefined%
  \global\let\svgscale\undefined%
  \makeatother%
  \begin{picture}(1,0.5488907)%
    \lineheight{1}%
    \setlength\tabcolsep{0pt}%
    \put(0,0){\includegraphics[width=\unitlength,page=1]{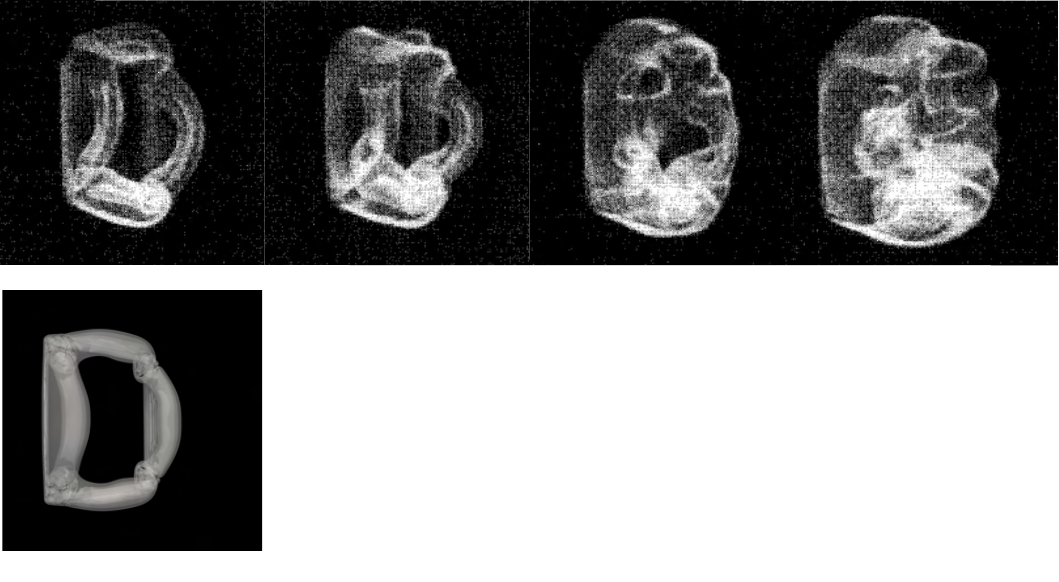}}%
    \put(0.03650099,0.00231317){\color[rgb]{0,0,0}\makebox(0,0)[lt]{\lineheight{1.25}\smash{\begin{tabular}[t]{l}\textbf{$a)\,\, t^*=2.0$}\end{tabular}}}}%
    \put(0.28641809,0.00231317){\color[rgb]{0,0,0}\makebox(0,0)[lt]{\lineheight{1.25}\smash{\begin{tabular}[t]{l}\textbf{$b)\,\, t^*=2.5$}\end{tabular}}}}%
    \put(0.53633516,0.00231317){\color[rgb]{0,0,0}\makebox(0,0)[lt]{\lineheight{1.25}\smash{\begin{tabular}[t]{l}\textbf{$c)\,\, t^*=3.0$}\end{tabular}}}}%
    \put(0.7863441,0.00231317){\color[rgb]{0,0,0}\makebox(0,0)[lt]{\lineheight{1.25}\smash{\begin{tabular}[t]{l}\textbf{$d)\,\, t^*=4.0$}\end{tabular}}}}%
    \put(0,0){\includegraphics[width=\unitlength,page=2]{fig/vortex_square_plate_oblique.pdf}}%
  \end{picture}%
\endgroup%

\end{graphicalabstract}

\begin{highlights}
\item Biot-Savart boundary conditions produce accurate unbounded flow simulations on Eulerian domains minimally larger than the immersed solid geometries.
\item Iterative partitioned solution of the non-locally coupled immersed surfaces and domain surfaces in the projection step requires minimal adjustment to classic velocity-pressure incompressible Navier-Stokes solvers.
\item Simplified oct-tree clustering for the Fast Multi-level Method (FM$\ell$M) integral evaluation has bounded error, optimal $O(N)$ operations, and efficient parallel execution on GPUs.
\end{highlights}


\end{frontmatter}



\section{Introduction}
When simulating external viscous flow problems with an Eulerian method using velocity and pressure as the primary variables, the ideally unbounded domain must be truncated to a finite size and artificial domain boundary conditions must be imposed on the computational domain's exterior. These typically consist of a known velocity normal to the boundary and a known normal pressure gradient \cite{Gresho1987}. These can lead to significant \emph{blockage} errors when the computational domain is not \emph{sufficiently large}, restricting the decay of the potential flow induced by the body, or advecting/diffusing vorticity through the domain's boundary \cite{Colonius2008}. While domain sensitivity study can determine the extent of the blockage, there is no \emph{a priori} rule for estimating their magnitude, and it is computationally inefficient to add grid cells to fill a large domain.

To reduce blockage effects and computational cost, domain augmentation method use a large coarse grid to provide artificial boundary conditions to a small well-resolved domain \cite{Colonius2008}. The outer flow field can also be augmented with a vortex particle-mesh method, reducing dispersive errors \cite{Billuart2023AFlows}. Such methods enable tracking the wake for an extended time while enabling accurate modeling of the developing boundary layer on the object's surface and can also include immersed boundaries \cite{Marichal2015UnboundedMethods}. However, these methods rely on a coupling of an external Lagrangian solver to an internal Eulerian solver formulated in different variables.

An alternative approach is to utilize some form of potential flow boundary conditions, following the classical \emph{Helmholtz} decomposition of the velocity field $\vec u(\vec x)$ into an irrotational and an incompressible part
\begin{equation}\label{eq:u_vort}
    \vec{u} = \vec{u}_\omega + \nabla\phi + \vec{U}_\infty = -\nabla^{-2}\left(\nabla \times \vec{\omega}\right) + \nabla\phi + \vec{U}_\infty
\end{equation}
where $\vec{u}_\omega$, $\nabla\phi$, $\vec{U}_\infty$ are the incompressible, irrotational, and the free-stream components, respectively. While these methods are elegant, they rely either on a single unbounded direction as in \cite{Grosch1977NumericalTransforms} or formulating the problem in a coordinate system where two unbounded directions are transformed to a single one (for example, from Cartesian to Cylindrical in \cite{Levy2022SolvingMethod}), which do not occur regularly in practical flow applications.

A related class of methods solves unbounded Poisson equations by incorporating the unbounded conditions as Dirichlet boundary data, either by successive relaxed iterations of the Poisson problem with the Dirichlet data computed using the previous solution \cite{Miller2008AnBoundaries} or by using the James-Lackner \cite{Jamb1977TheDistributions, Lackner1976COMPUTATIONEQUILIBRIA} algorithm that solves two nested domain Poisson problems, the inner domain with homogeneous Neumann boundary conditions and outer domain with inhomogeneous Neumann boundary conditions computed from a boundary convolution of the inner solution with the fundamental solution to the Poisson operator. To accelerate the convolution, multilevel approximation \cite{Mccorquodale2006ADimensions} or local correction methods \cite{Kavouklis2019ComputationCorrections} can be used.

Finally, fundamental solutions to discrete operators (Lattice Green's functions or LGFs) and their asymptotic expansions can be used to solve the discrete Poisson equations on unbounded domains \cite{Liska2014AEquations, Beckers2022PlanarGrids}. 
These methods rely on the unbounded properties of the LGFs and can simulate unbounded viscous incompressible flow \cite{Liska2016ADomains} and flows with immersed-boundaries \cite{Liska2017AFunctions}. This allows for snug computational domains limited to regions of non-negligible vorticity or possibly a set of disjoint domains limited to active vorticity regions. Combined with a multi-resolution approach, the LGF approach can yield fast and accurate computation of 3D incompressible flows at a cost reduced compared to Cartesian meshes \cite{Dorschner2020AEquations}. However, the method relies on a structured grid and a constant coefficient Poisson equation, and the possibility of extending the method to unstructured mesh and variable coefficient Poisson equation problems (such as the one arising in \cite{Lauber2022}) is unclear.

In this manuscript, we propose a novel method for imposing unbounded external flow boundary conditions on a finite Eulerian domain via a Biot-Savart integral solution applied on the domain's boundary.
The method retains the primitive variables formulation ($\vec u$ and the pressure $p$) of the governing equations, is compatible with immersed boundary methods, variable coefficient Poisson equations, fast pressure projection methods, and does not require additional meshes to describe the flow. In the next section, we describe the Biot-Savart boundary conditions and how they are injected into a projection algorithm to solve the \emph{Navier-Stokes} equations. Next, we demonstrate the insensitivity of this unique approach to errors in the velocity reconstruction on the boundary, and use that to develop a simplified multi-level clustering approach for an optimal $O(N)$ Fast Multi-level Method to evaluate the Biot-Savart integral. We validate the correctness of the coupled equations using accelerating flow examples where the external flow is truly irrotational and demonstrate that the new boundary conditions vastly outperform standard reflective boundary conditions on significantly smaller domains. We present an unexpected result: the new method continues to perform well when the fundamental assumption of potential flow external to the domain is violated, and we use the unstable deflected wakes generated by fast flapping foils to study the limits of the approach. We finish this manuscript by discussing the method's applicability to various flows in science and engineering.

\section{Method}

\begin{figure}
    \centering
    \def\svgwidth{0.5\columnwidth}
\begingroup%
  \makeatletter%
  \providecommand\color[2][]{%
    \errmessage{(Inkscape) Color is used for the text in Inkscape, but the package 'color.sty' is not loaded}%
    \renewcommand\color[2][]{}%
  }%
  \providecommand\transparent[1]{%
    \errmessage{(Inkscape) Transparency is used (non-zero) for the text in Inkscape, but the package 'transparent.sty' is not loaded}%
    \renewcommand\transparent[1]{}%
  }%
  \providecommand\rotatebox[2]{#2}%
  \newcommand*\fsize{\dimexpr\f@size pt\relax}%
  \newcommand*\lineheight[1]{\fontsize{\fsize}{#1\fsize}\selectfont}%
  \ifx\svgwidth\undefined%
    \setlength{\unitlength}{679.72156915bp}%
    \ifx\svgscale\undefined%
      \relax%
    \else%
      \setlength{\unitlength}{\unitlength * \real{\svgscale}}%
    \fi%
  \else%
    \setlength{\unitlength}{\svgwidth}%
  \fi%
  \global\let\svgwidth\undefined%
  \global\let\svgscale\undefined%
  \makeatother%
  \begin{picture}(1,0.47989949)%
    \lineheight{1}%
    \setlength\tabcolsep{0pt}%
    \put(0,0){\includegraphics[width=\unitlength,page=1]{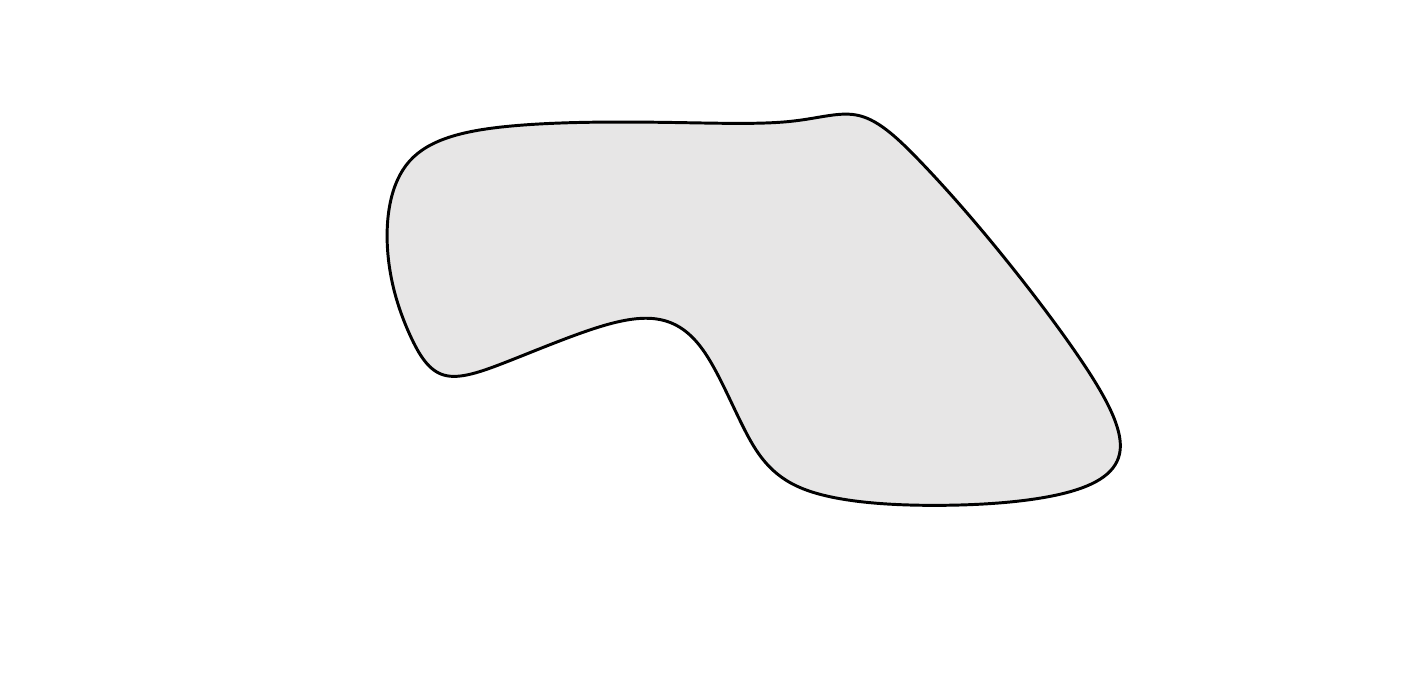}}%
    \put(0.48559941,0.30569118){\makebox(0,0)[lt]{\lineheight{1.25}\smash{\begin{tabular}[t]{l}$\mathcal{B}$\end{tabular}}}}%
    \put(0.1593647,0.11521314){\makebox(0,0)[lt]{\lineheight{1.25}\smash{\begin{tabular}[t]{l}$\Omega$\end{tabular}}}}%
    \put(0,0){\includegraphics[width=\unitlength,page=2]{fig/domain.pdf}}%
    \put(0.14,0.32){\makebox(0,0)[lt]{\lineheight{1.25}\smash{\begin{tabular}[t]{l}$\partial\mathcal{B}$\end{tabular}}}}%
    \put(0.68193149,0.37079271){\makebox(0,0)[lt]{\lineheight{1.25}\smash{\begin{tabular}[t]{l}$\hat{n}$\end{tabular}}}}%
    \put(0,0){\includegraphics[width=\unitlength,page=3]{fig/domain.pdf}}%
    \put(0.39267693,0.045){\makebox(0,0)[lt]{\lineheight{1.25}\smash{\begin{tabular}[t]{l}$\partial\Omega$\end{tabular}}}}%
    \put(0,0){\includegraphics[width=\unitlength,page=4]{fig/domain.pdf}}%
  \end{picture}%
\endgroup%

    \caption{Schematic of the fluid domain $\Omega$ with an immersed body $\mathcal{B}$ and their common (wet) interface $\partial\mathcal{B}$. The computational domain $\Omega$ has external boundary $\partial\Omega$ with $\hat{n}$ the unit normal vector.}
    \label{Fig_1}
\end{figure}

\noindent The flow is governed by the incompressible \emph{Navier-Stokes} equations
\begin{align}
    \begin{array}{c}
     \rho\frac{\partial\vec{u}}{\partial t} +\rho\left(\vec{u}\cdot\nabla\right)\vec{u} = -\nabla p + \mu\nabla^2\vec{u}\\
    \nabla\cdot\vec{u}=0  
    \end{array}
 & \quad\quad\forall\ \vec{x}\in\Omega,
\end{align}
where $\rho,\mu$ are the fluid's density and viscosity and $\Omega$ is the fluid domain , see Fig.~\ref{Fig_1}. These equations are supplemented with the no-slip boundary condition on the immersed body $\cal B$ with boundary $\partial\cal{B}$
\begin{equation}
    \vec{u} = \vec{U}_b \quad \forall x \in \partial\cal{B}
\end{equation} 
and a velocity condition on the domain boundary $\partial\Omega$
\begin{align}
    \vec u(\vec x) = f(\vec x,\vec U_\infty) &\quad\forall\ \vec{x}\in\partial\Omega,
\end{align}
where $\vec{U}_b$ is the body velocity and $f$ is the imposed condition and $\vec U_\infty$ is the far-field velocity. This work focuses on simulating external flow, i.e. $f$ is the identity function applied on an infinite domain $\partial\Omega\rightarrow\infty$.

There are many possible conditions to apply on a finite domain to simulate the external conditions, and the simplest of these is a (local) reflection condition
\begin{align}\label{eq:BC_0}
    \frac{\partial \vec{u}_s}{\partial \hat{n}} = 0,\ \vec{u}_n = \vec U_\infty\cdot \hat n &\quad\forall\ \vec{x}\in\partial\Omega,
\end{align}
where we have decomposed the velocity field on the boundary in a boundary normal $\vec{u}_n$ and tangential part $\vec{u}_s$. We refer the reader to \cite{Gresho1987} for a discussion of pressure boundary conditions on the domain's exterior.  As mentioned previously, the boundary $\partial\Omega$ must be placed sufficiently far away from the body not to influence the decaying potential flow part of the velocity field and ensure enough space to properly advect the vorticity downstream. 

We propose to replace the typical ``local" domain boundary conditions such as Eq.~\ref{eq:BC_0} by a Biot-Savart integral over the vorticity inside the domain. A Green's function solution to the vorticity equation Eq.~\ref{eq:u_vort} gives the velocity at a point $\vec x$ as
\begin{equation}\label{eq:Biot}
    {\vec u}({\vec x}) = \vec U_\infty+f(\vec x; \vec\omega,\Omega) = \vec U_\infty+\int_\Omega K_{n}({\vec x} - \vec{y})\times \vec\omega({\vec y})\text{ d}\vec{y} \quad\quad \forall \vec x\, \in \partial\Omega
\end{equation}
where $K_n$ is the $n$-dimensional Biot-Savart kernel. This kernel takes the following form \cite{Eldredge2019MathematicalFlows}
\begin{equation}\label{eq:kernels}
    2D:\quad K(\vec r)\equiv -\frac{\vec{r}}{2\pi|\vec{r}|^2}, \qquad\qquad 3D:\quad K(\vec r)\equiv -\frac{\vec{r}}{4\pi|\vec{r}|^3}.
\end{equation}
The vortex-induced velocity of the Biot-Savart integral is simply added to the uniform background velocity $\vec U_\infty$ to determine the total velocity at any point. We will omit further discussion of the trivial uniform flow contribution in this section.

This formulation implicitly assumes the flow external to the domain is irrotational and that the flow is incompressible everywhere. This assumption is met when all the vorticity is confined within the computational domain such as for the flow around accelerated bodies when the time $\frac{tU}{L}\lesssim t^{\text{crit.}}$, with $t^{\text{crit.}}$ the critical time for vortex detachment \cite{Shusser2000EnergyRing}. However, we will show that the results of using Eq.~\ref{eq:Biot} in an Eulerian solver are still highly accurate for flows with wakes that extend to the domain boundary. 

Note that on a staggered grid, we only need to apply the integral equation (Eq.~\ref{eq:Biot}) to compute the velocity normal to the domain face. For ghost cells external to the computational domain, we use the local derivative conditions $\nabla\cdot\vec{u} = \nabla\times\vec{u}=0$ which are consistent with the assumption that the flow is potential outside the domain. Specifically, once we have used Eq.~\ref{eq:Biot} to set $u_n$ along the domain boundary we apply the zero curl condition to all tangential velocity components in the ghost cells and then the zero divergence condition to fill the remaining normal component each ghost cell. For example, in 2D we sequentially apply
\[ \frac{\partial \vec{u}_s}{\partial \hat{n}} = \frac{\partial \vec{u}_n}{\partial \hat{s}} \quad \text{and} \quad \frac{\partial \vec{u}_n}{\partial \hat{n}} = -\frac{\partial \vec{u}_s}{\partial \hat{s}}.\]
This process can continue for multiple layers of ghost cells, marching away from the domain boundary.

There are two primary issues with the use of Eq.~\ref{eq:Biot} as the domain boundary condition for the incompressible Eulerian solver: (i) projecting the velocity into a divergence-free field consistent with the Biot-Savart equation, and (ii) the fast evaluation of Eq.~\ref{eq:Biot}. We demonstrate in the following section that these issues may be efficiently and accurately overcome largely without modifying the method used to solve the Navier-Stokes equation.

\subsection{Biot-Savart Projection Algorithm}\label{sec:Biot_projection}

We use the standard projection scheme \cite{Chorin1967} to solve the coupled velocity-pressure system inside a predictor-correct method to achieve second-order temporal convergence of the flow variables \cite{Lauber2022}. 
To simplify the discussion, we present the method for the first step of the predictor-corrector; generalization to the second-order Heun's corrector or higher-order time integration method is straightforward. The projection method starts with an explicit estimate of the intermediate velocity field $u^*$
\begin{align}\label{eq:intermediate}
    \vec{u}^* = \vec{u}\,^t + \int_{t}^{t+\Delta t}\frac{1}{\text{Re}}\nabla^2\vec{u} -\left(\vec{u}\cdot\nabla\right)\vec{u}\text{ d}t &\quad\forall\ \vec{x}\in\Omega (N),
\end{align}
where $\text{Re}$ is the Reynolds number of the flow. Depending on the choice of integration method for the right-hand-side, we obtain explicit or semi-implicit methods. Here we will focus on fully explicit methods. 



Projection of the divergent velocity field onto the solenoidal space is achieved by computing a pressure field $p$ that removes the divergent part of the intermediate velocity field
\begin{align}\label{eq:poisson}
      &\nabla\cdot\frac{\Delta t}{\rho}\nabla p = \nabla\cdot \vec{u}^* &\quad\forall\ \vec{x}\in\Omega (N),\\
      \label{eq:project}
      &\vec{u}^{t+\Delta t} = \vec{u}^*-\frac{\Delta t}{\rho}\nabla p &\quad\forall\ \vec{x}\in\Omega (N),
\end{align}
also subject to the standard boundary condition
\begin{align}\label{eq:BC_2}
      &\frac{\partial \vec{u}^{t+\Delta t}_s}{\partial \hat{n}} = 0,\ \vec{u}^{t+\Delta t}_n = U_n &\quad\forall\ \vec{x}\in\partial\Omega (N^{\cal S}).
\end{align}

The main contribution of this manuscript is injecting the Biot-Savart integral into the classical projection scheme and solving the resulting coupled problem. The momentum update using the new Biot-Savart domain condition proceeds as before: forming the intermediate velocity with Eq.~\ref{eq:intermediate} and projecting it to be divergence free with Eq.~\ref{eq:project}. However, the new boundary condition introduces a coupling between the body and domain boundaries. To see this, we substitute the update equation for the velocity at the new time-step (Eq.~\ref{eq:project}) into the Biot-Savart equation for the boundary velocity (Eq.~\ref{eq:Biot}) to give to complete equation for the boundary velocity at $t+\Delta t$
\begin{align}\label{eq:biot-press}
    &\vec{u}^{t+\Delta t}(\vec x_b) = f(\vec{x_b},\nabla\times \vec{u}^{t+\Delta t},\Omega) = f(\vec{x_b},\nabla\times\vec{u}^*,\Omega)-f\left(\vec{x_b},\nabla\times\frac{\Delta t}{\rho}\nabla p,\Omega\right) 
\end{align}
While the gradient of a field cannot induce curl in an \textit{unbounded} fluid region ($\nabla\times\nabla\phi\equiv 0$), the pressure generates a thin layer of vorticity \textit{on the body boundary} $\partial\cal B$ every time-step \cite{Morton1984GeophysicalTy}. This means that the last term in Eq.~\ref{eq:biot-press} is non-zero, introducing a non-local coupling between every point on the body boundary and every point on the domain boundary.

Because of the pressure contribution in Eq.~\ref{eq:biot-press}, the pressure projection (Eq.~\ref{eq:poisson}) becomes
\begin{equation}\label{eq:complex_poisson}
    \nabla\cdot\left[\frac{\Delta t}{\rho}\nabla p + f\left(\vec{x_b},\nabla\times\frac{\Delta t}{\rho}\nabla p,\Omega\right)\right] = \nabla\cdot\left[ \vec{u}^* + f(\vec{x_b},\nabla\times\vec{u}^*,\Omega)\right], \quad\forall\
    \begin{array}{l}
    \vec{x} \in \Omega(N),\\
    \vec{x}_b \in \partial\Omega(N^\mathcal{S})
    \end{array}
\end{equation}
The linear operator resulting from discretizing Eq.~\ref{eq:poisson} with a local finite-difference finite-volume or finite-element method has only $O(N)$ non-zero terms, but the coupling induced by the Biot-Savart equation results in a dense matrix with at least $O(N^{1+S})$ non-zero terms. Note that this coupling is similar to the additional terms present in the Poisson equation when correctly imposing pressure boundary conditions for immersed boundary methods; see, for example, \cite{Taira2007, Lauber2022}. Failing to correctly enforce the boundary conditions during projection leads to large errors in the predicted pressure and forces.

We must accurately include the coupling term in Eq.~\ref{eq:biot-press}, but we want to avoid constructing the dense matrix or requiring substantial changes to existing Poisson solvers to make it practical for Eulerian schemes to use Biot-Savart boundary conditions. This is possible through a iterative operator-split method akin to the James-Lackner algorithm for potential flows
\begin{equation}\label{eq:differed_poisson}
    \nabla\cdot\frac{\Delta t}{\rho}\nabla p^{k+1} = \nabla\cdot\left[ \vec{u}^* + f\left(\vec{x_b},\nabla\times\left[\vec{u}^*-\frac{\Delta t}{\rho}\nabla p^k\right],\Omega\right)\right], \quad\forall\
    \begin{array}{l}
    \vec{x} \in \Omega(N),\\
    \vec{x}_b \in \partial\Omega(N^\mathcal{S}),
    \end{array}
\end{equation}
using the previous pressure solution $p^k$ to treat the Biot-Savart term as a residual and iterating until both the incompressible and boundary conditions are satisfied. Eulerian solvers typically use iterative methods to determine the pressure field in the projection step, making this a natural way to incorporate the coupling between the domain and body boundaries, without modifying the Poisson solver itself. In this work, we use a multigrid solver for the Poisson system, as described in \cite{Weymouth2022Data-drivenProjection}, and updating the domain velocities and corresponding Poisson residuals at the end of each V-cycle is sufficient to ensure that the final pressure and velocity field without increasing the number of iterations.

Our integration of Eq.~\ref{eq:differed_poisson} into the unsteady momentum update significantly improves the stability of the coupled system compared to similar applications in potential flow such as ~\cite{Miller2008AnBoundaries}. The error introduced by the operator partitioning is bounded by the change in pressure on the body over a single time step. We therefore never require relaxation to achieve convergence, even when impulsively starting a flow from rest with domain boundaries a small fraction of the body length away from the body boundaries. 
In addition, the resulting coupled system is remarkably robust to errors in the velocity reconstruction because the Biot-Savart equation is only used to set the boundary conditions, not update the entire flow field as in vortex-based methods. Consider the effect of neglecting the contribution from a vorticity source $\Delta\omega$ that has just left the domain, meaning this contribution is missing from $\nabla\times\vec u^*$ in Eqs.\ref{eq:biot-press} and \ref{eq:differed_poisson}. As only the boundary velocities are updated in Eq.~\ref{eq:biot-press}, the momentum errors in the local Eulerian Navier-Stokes update are limited to the downwind characteristics from the domain boundary, and the impact on the (critical) inflow boundary regions are typically the smallest, as they are the farthest from the vorticity leaving the outflow boundary regions. 


\subsection{Fast evaluation of the Biot-Savart integral}

The aim of Biot-Savart boundary conditions is to reduce the computational cost compared to a large grid by providing an accurate estimate of external flows using small domains. While the boundary conditions are agnostic of the method used to compute the integral in Eq.~\ref{eq:Biot}, a naive convolution over $N$ vorticity ``source'' locations $\vec y$ would require $O(N)$ operations for \textit{every} ``target'' velocity $\vec u(\vec x)$ on the boundary. The overall $O(N^{1+\cal S})$ cost would likely overwhelm any benefit of the reduced grid size, motivating a faster Biot-Savart evaluation approach.


Highly precise Fast Multipole Methods (FMMs) have been developed to a reduce this cost to $O(\log N)$ or even $O(1)$ operations per target, \cite{Greencard1987ASimulations,Ying2004ADimensions, Fong2009TheMethod}. Our context allows for significant specialization and simplification to a standard FMM. (i) A general FMM considers a general set of source locations, and they are typically optimized to evaluate on the same set of targets. In contrast, our sources are equally spaced over the domain and our targets are equally spaced on it's boundary, meaning we have only $O(N^{\cal S})$ targets which are exclusively exterior to the source set. (ii) The high error tolerance discussed in the section above means truncating the multipole has no effect on the satisfaction of the divergence-free constraint (set by the Poisson solver tolerance) and little effect on the solution accuracy overall.

\begin{figure}
    \centering
    \def\svgwidth{0.7\columnwidth}
    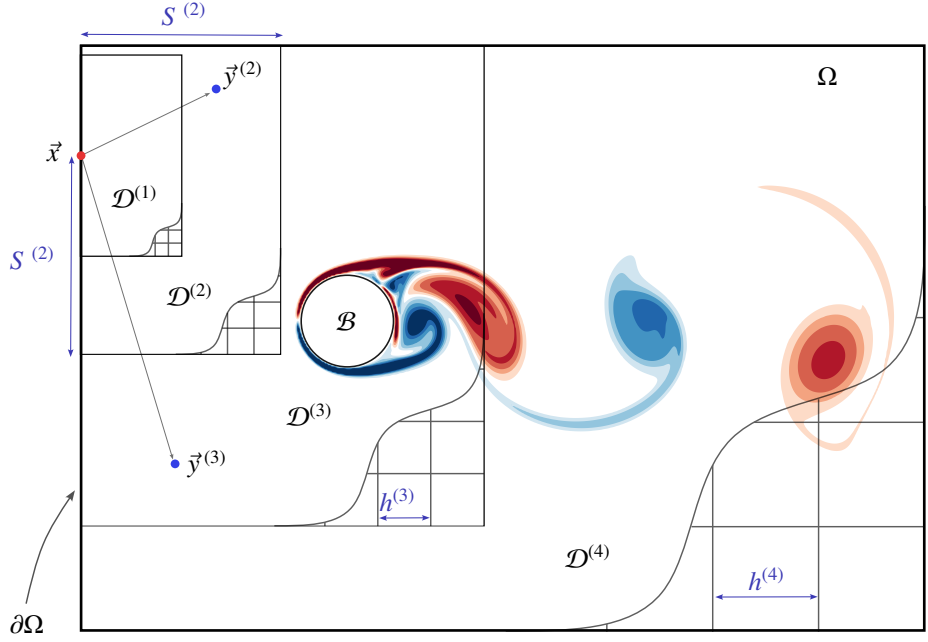
    \caption{Schematic of the multilevel approach to evaluate the Biot-Savart integral for a boundary point $\vec x$ over the nested subdomains $\mathcal{D}^{(1)} \cup \mathcal{D}^{(2)} \cup \mathcal{D}^{(3)} \cup \mathcal{D}^{(4)}$. The second domain half-width is shown, ${S}^{(2)}$. The computational domain $\Omega$ and its boundary $\partial\Omega$ and the immersed body $\mathcal{B}$ are shown. We also show a schematic of the coarsening of the grid between the domains and the multilevel pooling $\mathcal{P}^{(3)\to(4)}$ of the cell's circulation (circular-colored markers). 
    }
    \label{Fig_2}
\end{figure}

Based on this context, we use a simple multi-level vorticity source clustering method which leverages their uniform distribution in our Eulerian domain. We construct a multi-level vorticity field $\omega^{(i)}$ where level $i=1$ is the finest and $l$ is the coarsest and the cell size doubles with each level $h^{(i+1)}=2h^{(i)}$. The levels are filled from finest to coarsest recursively by uniformly pooling the circulation $\Gamma=\omega h^2$ on the level above 
\begin{equation}
\Gamma^{(i+1)}=\mathcal{P}^{(i)\to(i+1)}\Gamma^{(i)}
\end{equation}
where the pooling $\mathcal{P}$ for a cell at level $i+1$ simply sums over the corresponding sub-cells in level $i$, see Fig.~\ref{Fig_2}. Thus, our multi-level field conserves the total circulation to machine precision, as recommended by \cite{Colonius2008} and others. Filling the entire multi-level source field requires $O(N)$ operations.

In this work, we present results from two boundary velocity reconstruction approaches using the multi-level source clustering described above:

\begin{enumerate}
    \item \textit{tree-sum}: In the first approach, the integral over the original domain $\Omega$ in Eq.~\ref{eq:Biot} is replaced with a sum over the contribution from a sequence of subdomains
    \begin{equation}\label{eq:multilevel}
        \Omega = \mathcal{D}^{(1)}(\vec x) \cup \mathcal{D}^{(2)}(\vec x) \cup \cdots \cup \mathcal{D}^{(l-1)}(\vec x) \cup \mathcal{D}^{(l)}(\vec x),
    \end{equation}
    where we define the sub-domains as nested rectangular boxes of half-width $S^{(i)}=h^{(i)}\tilde S$ centered on the target location $\vec x$, Fig \ref{Fig_2}. As the sub-domains are non-overlapping and fully cover $\Omega$, 
    Eq.~\ref{eq:Biot} becomes 
    \begin{equation}\label{eq:biot_sum}
        \vec u(\vec{x}) \approx \sum_{i=1}^{l} f(\vec{x}; \vec \omega^{(i)},\mathcal{D}^{(i)}(\vec x)).
    \end{equation}
    This corresponds to a direct sum for each target $\vec x$ over an oct-tree source field in 3D (or a quad-tree field in 2D). Therefore, we call this the ``tree-sum'' approach. As each subdomain has $O(\tilde S^n)$ sources and the number of levels $l=O(\log N)$, the tree-sum requires $O(\tilde S^n\log N)$ operations per target.
    \item \textit{FM$\ell$M}: In the second approach, we \textit{also} cluster the domain boundary target points $\vec  x^{\,(i)}$, doubling the spacing at each level. We then define the level-$i$ kernel interaction velocities as
    \begin{equation}\label{eq:interaction}
        \vec v^{\,(i)}(\vec x^{\,(i)}) = f(\vec x^{\,(i)}; \vec \omega^{(i)},\mathcal{D}^{(i)}(\vec x^{\,(i)})).
    \end{equation}
    The target velocities are then unpooled recursively from the coarsest level to the finest
    \begin{equation}\label{eq:unpool}
        \vec u^{\,(i)} \approx \vec v^{\,(i)}+\mathcal{P}^{(i+1)\to(i)} \vec u^{\,(i+1)}
    \end{equation}
    where $u^{\,(l+1)}=0$ and the unpooling operator is bi-linear interpolation to the target locations $\vec x^{\,(i)}$. As the interaction velocities are independent at each level, they require a constant $O(\tilde S^n)$ operations per target. Therefore, we call this the Fast Multi-\textit{level} Method (FM$\ell$M). 
\end{enumerate}

A key difference between these approaches and a classic FMM is that the contributions from each clustered source point are obtained by evaluating Eq.~\ref{eq:kernels} directly and not through a classical or kernel-independent multipole expansion, as done in \cite{Ying2004ADimensions, Liska2014AEquations}. As such, the domain half-width $\tilde S$ controls the reconstruction error and \ref{A0} proves that this error is bounded and inversely proportional to the number of subdomain points $\tilde S^n$. The results sections below demonstrate that these approaches are more than sufficient for high-accuracy external flow predictions.

Finally, we note that we can parallelize the velocity reconstruction to further accelerate the method since each Biot-Savart kernel evaluation is independent. Our implementation applies GPU or CPU multi-threading to parallelize the calculation and pooling of the vorticity across the sources, as well as the kernel function evaluation and velocity unpooling across the targets.

\section{Confined vorticity applications}

We first validate the application of the Biot-Savart boundary conditions using a set of impulsive flow problems. The flow is truly irrotational outside the computational domain, matching the implicit assumptions required to use the Biot-Savart equations to reconstruct the velocity on the boundary. 

Our Biot-Savart boundary conditions are implemented inside \textbf{\color{cyan}{{WaterLily.jl}}}\footnote{\url{https://github.com/WaterLily-jl/WaterLily.jl}}, a fast, immersed-boundary flow solver based on the Boundary data immersion method \cite{Maertens2015} that can execute both on CPU and GPU \cite{Weymouth2023WaterLily.jl:Execution}. To leverage the GPU capability of the flow solver and fully benefit from our novel boundary conditions, we perform all computations presented herein on an NVIDIA-A6000 GPU (48GB memory). The complete code and instructions needed to reproduce these results are freely available at \url{https://github.com/weymouth/BiotSavartBCs.jl}.

\subsection{Flow around a 2D cylinder at $\text{Re}=550$}

\begin{figure}
    \centering
    \begin{subfigure}{.4\textwidth}
        \hspace{-5mm}
        \includegraphics[width=\textwidth]{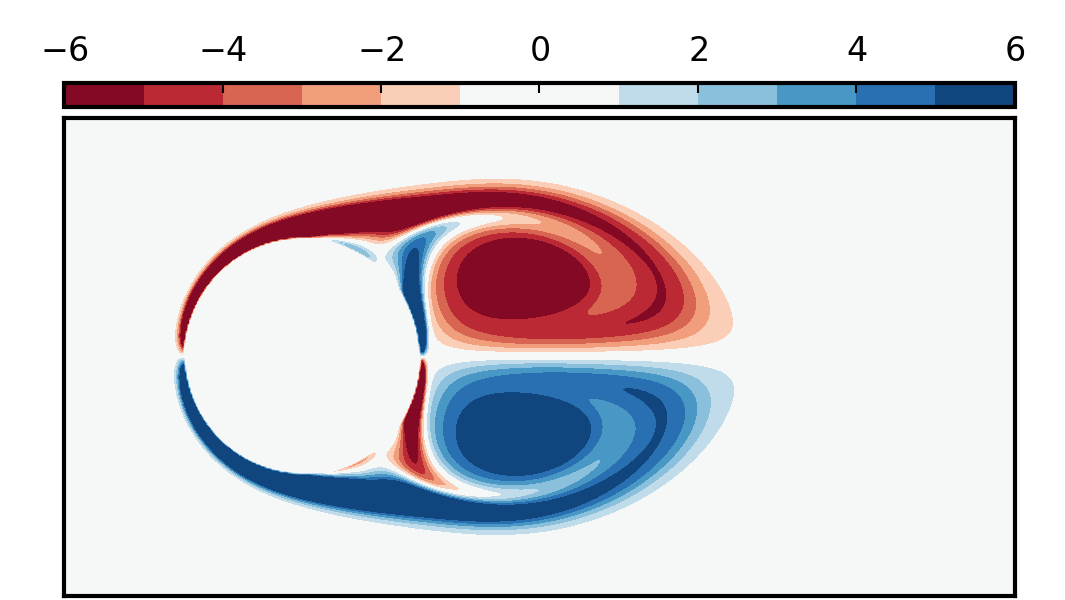}
        \vspace{10mm}
    \end{subfigure}%
    \begin{subfigure}{.55\textwidth}
        \centering
        \includegraphics[width=\textwidth]{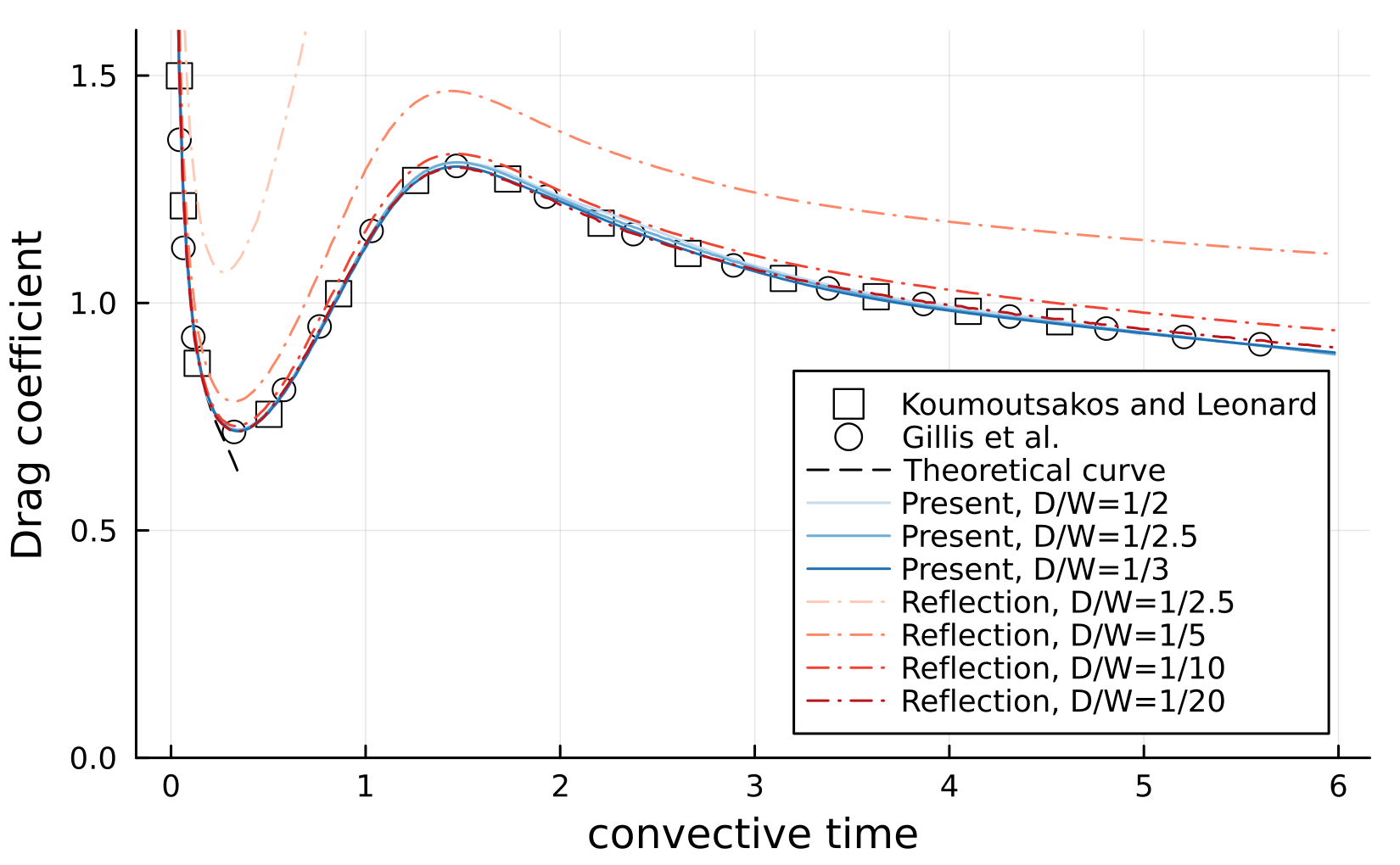}
        \vspace{-5mm}
    \end{subfigure}%
    \caption{Impulsively started circular cylinder at $\text{Re}=550$ validation case. (Right) Vorticity field $\omega D/U$ at convective time $tU/D=4$ over the full $2D\times 4D$ domain. (left) Early time history of the drag coefficient $C_d=F_d/\frac 12 \rho U^2 D$ for various domain blockage ratios $D/W$. The present Biot-Savart BCs match the theoretical and vortex-based methods even with 50\% blockage.
    }
    \label{fig:cylinder_force}
\end{figure}

We start our analysis with the flow behind a 2D circular cylinder of diameter $D$. The cylinder is immersed in a viscous fluid and impulsively accelerated to free-stream velocity $U$. The viscosity is set such that the Reynolds number is $\text{Re}=\frac{UD}{\nu}=550$. We set $D=128$ cells for all tests and use a $2\times 1$ rectangular domain, varying the width $W$ from $=2D\ldots 24D$. Figure~\ref{fig:cylinder_force}(left) shows the vorticity field on the smallest domain at convective time $t^*=tU/D=4$.

Figure~\ref{fig:cylinder_force}(right) compares the predicted drag coefficient using the new Biot-Savart boundary conditions to the analytical solution \cite{Collins1973TheCylinder} (for $t^*\ll 1$) and two sets of reference data from high-resolution vortex-based methods \cite{KoumoutsakostAN1995High-resolutionMethods, Gillis2019AMethod}. The new condition is able to accurately predict the total drag force on the cylinder, matching the theoretical and vortex-based methods even when using the smallest $4D\times2D$ domain with 50\% blockage. 
Reflective boundary conditions induce extremely large blockage effects for small domains and only slowly converge to the external flow solution. Even using a domain 100 times larger produces noticeable deviations for $t^*>3.5$. 


\subsection{2D and 3D flow around an accelerating disk}

The flow around a 2D or 3D disk accelerated from rest in a quiescent flow is purely potential for $t\to 0^+$. As a result, an analytical expression for the added mass coefficient can be used to further validate our novel potential flow boundary condition when injected in a moment step of the \emph{Navier-Stokes} equations. 

The disk of radius $R$ and diameter $D=2R$ is given a constant acceleration $a$ for the initial non-dimensional time $t^*=at/U\le 1$ and is given a constant speed after this, ie
\begin{equation}
    U(t) = \begin{cases}
        at, \quad \text{if } at/U\le1,\\
        U, \quad \text{else}.
    \end{cases}
\end{equation}
We perform 2D and 3D simulations of the flow around a plate/circular disk immersed in a domain of dimension $1.5D\times1.5D\,\,(\times1.5D)$ with a resolution of 86 cells per diameter ($D$). From potential flow theory, the added-mass coefficient for an impulsively started flow is
\begin{equation}
    Ca  = \frac{F_{am}}{\rho D^3 a} = \frac{F_{am}}{\rho D^2U^2}\left(\frac{1}{a^*}\right) 
\end{equation}
where $Ca$ is the classical added-mass coefficient. We simulate the 2D or 3D flow until $t^*=3$. Once the disk reaches its peak velocity, the Reynolds number is $\text{Re}=\frac{UR}{\nu}=1000$.

We note here that the 2D and 3D problems possess axes of symmetry that would allow only a portion of the domain to be modeled. This can be easily accounted for in the Biot-Savart boundary conditions by adding the contributions of the images of each vortex cell, see \ref{A1}. However, to avoid unnecessary complications in this manuscript, we model the full plate and do not use the problem's symmetries.

\begin{figure}
    \centering
    \includegraphics[trim={0cm 1cm 0cm 0.5cm},clip,width=0.8\textwidth]{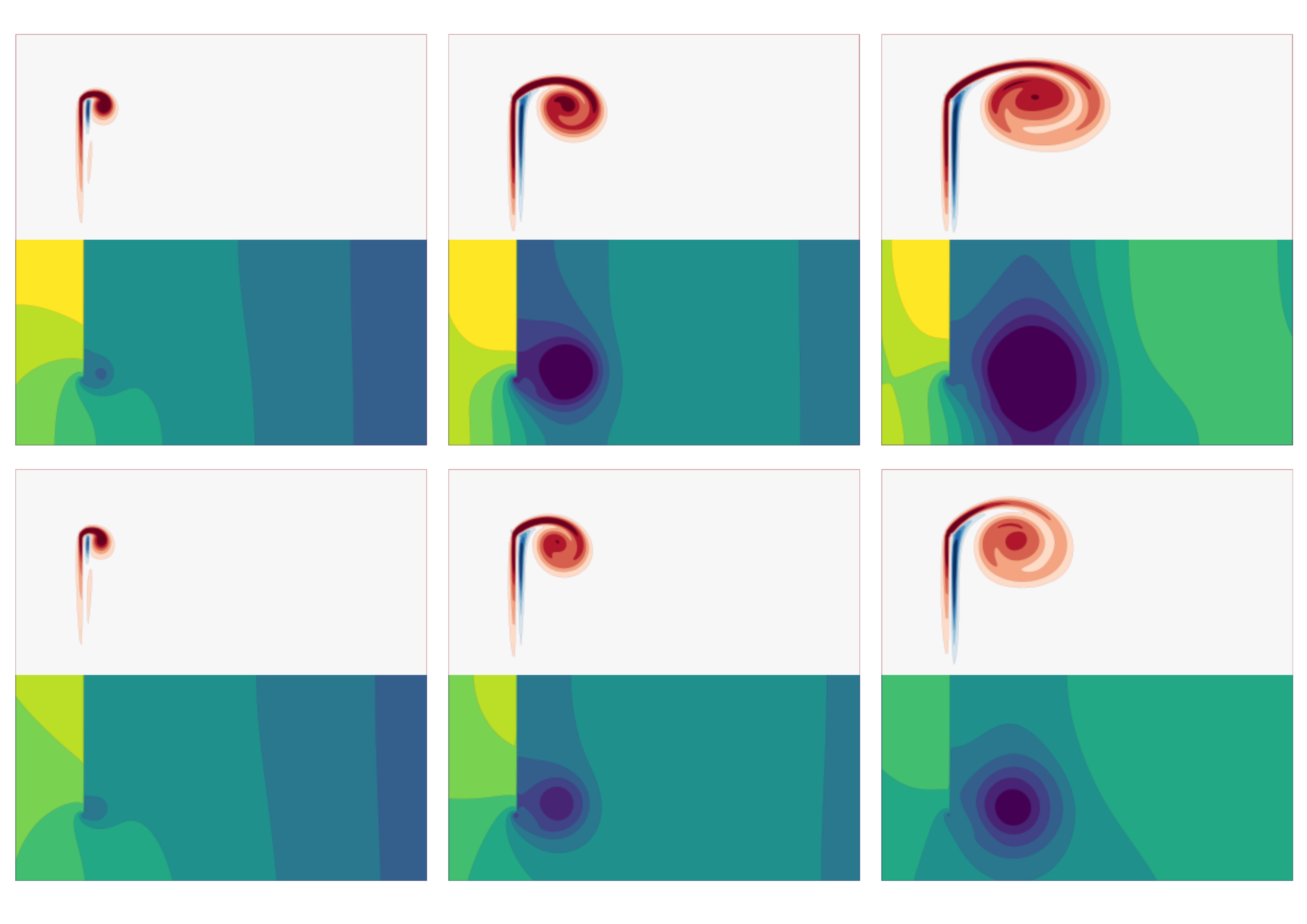}
    \caption{Flow around an initially stationary disk accelerated in a quiescent flow at three convective times, $t^*\in [1,2,3]$, using the reflective boundary conditions (top) and the Biot-Savart boundary condition (bottom). A slice through the middle of the entire computational domain is shown. The top half shows 10 iso-contour of the vorticity, equally spaced in the interval $\omega R/U\in\pm20$. The bottom half shows the pressure field, with 10 iso-contour equally spaced in $p/\rho U^2\in\pm2$.}
    \label{fig:disk_flow_1}
\end{figure}

Figure~\ref{fig:disk_flow_1} shows the vorticity and pressure field for three consecutive times during the disk's motion on the midplane of the 3D circular disk. The top row shows the results with the standard reflective boundary conditions (the entire computational domain is shown), and the bottom row shows the flow obtained with the Biot-Savart boundary conditions. High blockage effects are apparent for the reflective boundary conditions, which generate a stronger shear layer that feeds a stronger vortex located further downstream than the Biot-Savart results. The error in the pressure field due to the homogeneous Neumann condition placed close to the immersed body generates pressure peaks twice as large as with the Biot-Savart boundary conditions.

\begin{figure}
    \centering
    \begin{subfigure}{.5\textwidth}
        \centering
         \includegraphics[width=\textwidth]{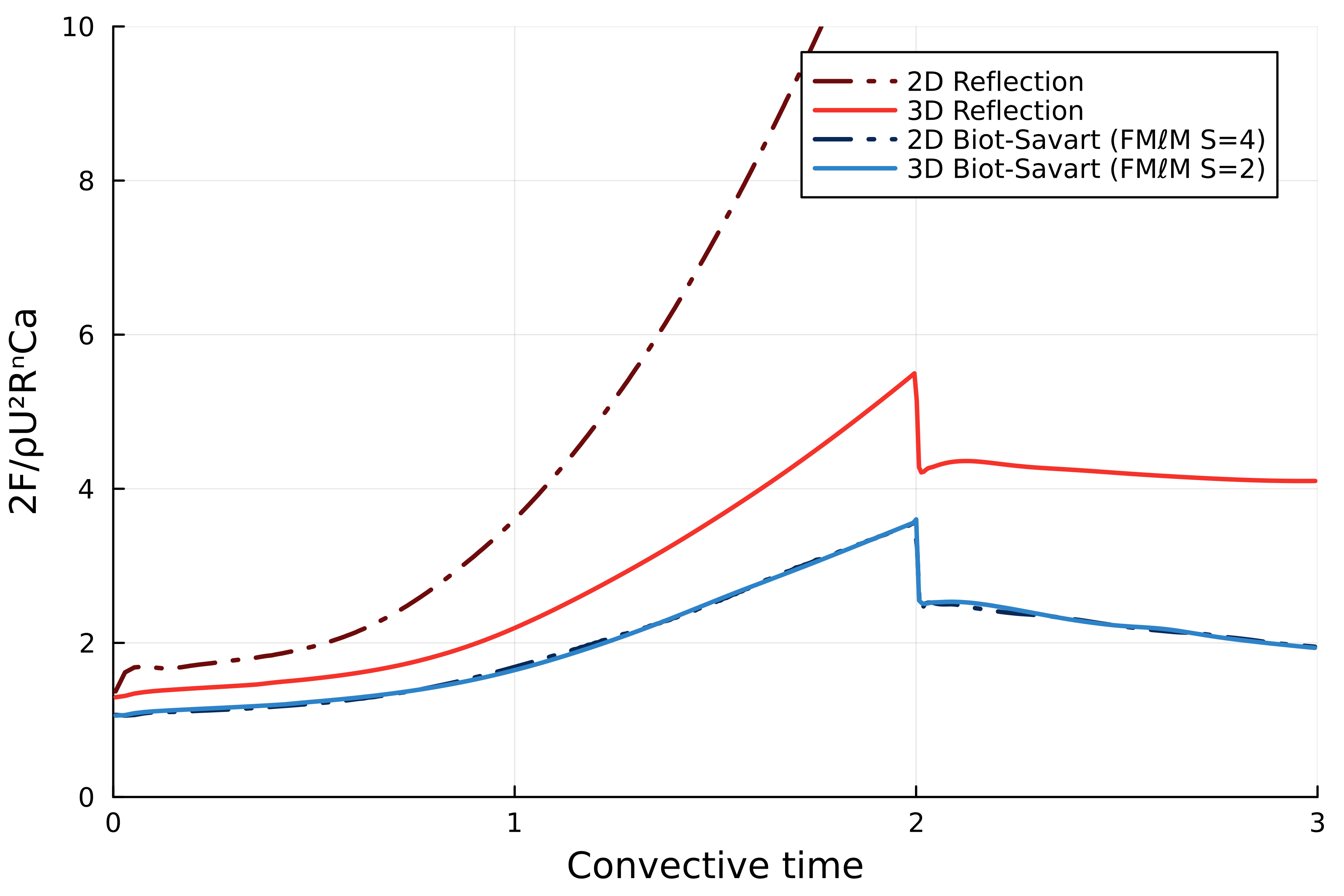}
         \caption{}
    \end{subfigure}%
    \begin{subfigure}{.5\textwidth}
        \centering
        \includegraphics[width=\linewidth]{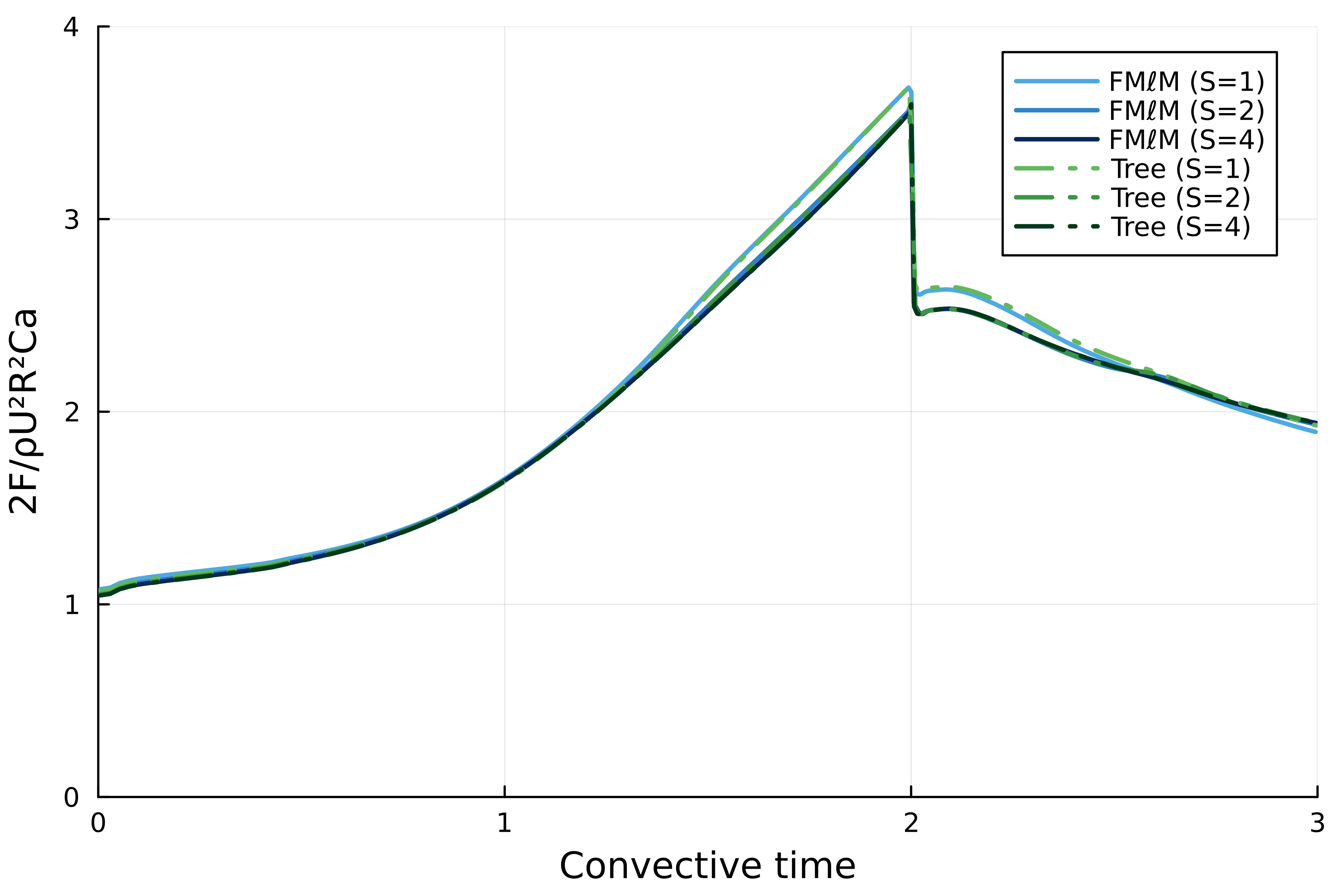}
        \caption{}
    \end{subfigure}%
    \caption{Instantaneous pressure forces normalized by added mass acting on the 2 and 3D disk accelerated from rest for (a) the two different boundary conditions and (b) for different Biot-Savart kernels (tree-sum and FM$\ell$M) and different subdomain half-width $\tilde S$ for the 3D disk. The convective time corresponds to the snapshots shown in Fig.~\ref{fig:disk_flow_1}.}
    \label{fig:disk_forces}
\end{figure}

Fig.~\ref{fig:disk_forces} shows the time trace of the pressure force acting on the accelerated disk, as well as the equivalent 2D simulation on an accelerating flat plate for comparison. The impact of the error in in the pressure field is apparent in the difference in the forces. Here, the Biot-Savart boundary condition can exactly recover the added-mass force on the first time step, while the reflective boundary condition overestimates it. For larger $t^*$, the effect of the boundary condition is still observable; the reflective boundary condition significantly overestimates the pressure forces for all $t$. 

Next we demonstrate the low sensitivity of the Eulerian flow solver using the simple partitioned approach of Eq.~\ref{eq:differed_poisson} to the implementation details of the Biot-Savart kernel summation. Fig.~\ref{fig:disk_forces}(b) compares the instantaneous drag force acting on the 3D disk using the tree-sum and FM$\ell$M approach over a range of clustering subdomain sizes. The results are extremely consistent, with any $\tilde S \ge 2$ predicting identical forces and only the smallest possible half-width $\tilde S=1$ leading to a few percent increase in the predicted peak force at $t^*=2$. In addition, the forces are identical using the two approaches despite the additional unpooling step of the FM$\ell$M.

This minimal sensitivity means that we can maintain accuracy while maximizing the reconstruction speed. For the remainder of the manuscript we will use the FM$\ell$M approach with $\tilde S = 2$ for 3D flows, and $\tilde S = 4$ for 2D. We therefore use $O(N)$ operations for clustering the sources and $O(N^{\cal S})$ operations for reconstructing the target velocities for each V-cycle of the pressure solver, which itself is $O(N)$. In practice, we find that using Biot-Savart boundary conditions on small domains in 2D can double the simulation speed compared to using reflection condition on the same domain because the artificially fast flow speeds shown in Fig.~\ref{fig:disk_flow_1} decrease the possible time-step. In 3D we find simulation times are basically identical using Biot-Savart boundary conditions or reflection conditions on the same domain, with the small reconstruction cost and increased time-steps roughly canceling.

\subsection{3D Flow around a square disk at $\text{Re}=125\,000$}

We next simulate the flow around a square plate of dimension $L\times L$, accelerated in the surface normal direction to a final Reynolds number $\text{Re}=125\,000$ to demonstrate the ability of our method to deal with highly separated and turbulent vortex formation. We focus on the initial vortex formation and position the plate at the center of a domain of size $2L\times2L\times2L$. The disk has a resolution of 384 cells per diameter ($D$), resulting in a $y^+\sim 32$. We simulate the flow evolution until $t^*<6$, where vortices typically stay attached to the body and only separate and travel downstream later. Simulations are performed on a single \texttt{NVIDIA-A100} PCIe card with 80GB (filled at $\sim80\%$ capacity) in a little less than 160 minutes for $1.8\times10^9$ DOF, resulting in execution time of 5.1~ns/DoF/dt.

\begin{figure}
    \centering
    \def\svgwidth{1\columnwidth}
\begingroup%
  \makeatletter%
  \providecommand\color[2][]{%
    \errmessage{(Inkscape) Color is used for the text in Inkscape, but the package 'color.sty' is not loaded}%
    \renewcommand\color[2][]{}%
  }%
  \providecommand\transparent[1]{%
    \errmessage{(Inkscape) Transparency is used (non-zero) for the text in Inkscape, but the package 'transparent.sty' is not loaded}%
    \renewcommand\transparent[1]{}%
  }%
  \providecommand\rotatebox[2]{#2}%
  \newcommand*\fsize{\dimexpr\f@size pt\relax}%
  \newcommand*\lineheight[1]{\fontsize{\fsize}{#1\fsize}\selectfont}%
  \ifx\svgwidth\undefined%
    \setlength{\unitlength}{508.29994502bp}%
    \ifx\svgscale\undefined%
      \relax%
    \else%
      \setlength{\unitlength}{\unitlength * \real{\svgscale}}%
    \fi%
  \else%
    \setlength{\unitlength}{\svgwidth}%
  \fi%
  \global\let\svgwidth\undefined%
  \global\let\svgscale\undefined%
  \makeatother%
  \begin{picture}(1,0.5488907)%
    \lineheight{1}%
    \setlength\tabcolsep{0pt}%
    \put(0,0){\includegraphics[width=\unitlength,page=1]{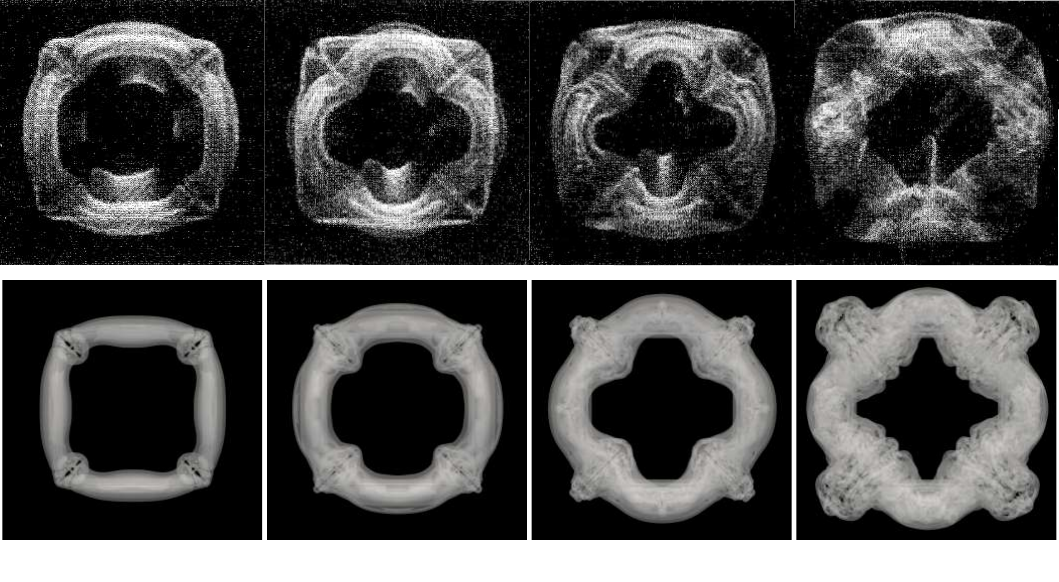}}%
    \put(0.03650099,0.00231317){\color[rgb]{0,0,0}\makebox(0,0)[lt]{\lineheight{1.25}\smash{\begin{tabular}[t]{l}\textbf{$a)\,\, t^*=2.0$}\end{tabular}}}}%
    \put(0.28641809,0.00231317){\color[rgb]{0,0,0}\makebox(0,0)[lt]{\lineheight{1.25}\smash{\begin{tabular}[t]{l}\textbf{$b)\,\, t^*=2.5$}\end{tabular}}}}%
    \put(0.53633516,0.00231317){\color[rgb]{0,0,0}\makebox(0,0)[lt]{\lineheight{1.25}\smash{\begin{tabular}[t]{l}\textbf{$c)\,\, t^*=3.0$}\end{tabular}}}}%
    \put(0.7863441,0.00231317){\color[rgb]{0,0,0}\makebox(0,0)[lt]{\lineheight{1.25}\smash{\begin{tabular}[t]{l}\textbf{$d)\,\, t^*=4.0$}\end{tabular}}}}%
  \end{picture}%
\endgroup%

    \caption{Vortex formation behind the square plate at $\text{Re}=125\,000$ at four different instants. The top row is the flow visualization from \cite{Higuchi1996Three-dimensionalPlates}, and the bottom row is the same view from our simulations using $\lambda_2L^2/U^2=-2\times10^{-7}$ for the vortex identification. The plate is omitted for clarity.}
    \label{fig:square_disk_comparison}
\end{figure}

We qualitatively compare the shape of the vortex ring forming behind our accelerating square plate with the flow visualization of the same cases in \cite{Higuchi1996Three-dimensionalPlates}, who used dye coloring to visualize the wake forming behind the accelerating plate. Results are presented in Figure~\ref{fig:square_disk_comparison_oblique}, \ref{fig:square_disk_comparison} for four instants during and after the acceleration phase. 

\begin{figure}
    \centering
    \def\svgwidth{1\columnwidth}
    
    \caption{Vortex formation behind the square plate at $\text{Re}=125\,000$ at four different instants viewed from an oblique angle. The top row is the flow visualization from \cite{Higuchi1996Three-dimensionalPlates}, and the bottom row is the same view from our simulations using $\lambda_2L^2/U^2=-2\times10^{-7}$ for the vortex identification. The plate is omitted for clarity. An animation of these still frames is provided in the supplementary material.}
    \label{fig:square_disk_comparison_oblique}
\end{figure}
An excellent agreement between the simulations and the experiments is found for this qualitative verification, both in the oblique and downstream view from the plate. The initial vortex roll-up starts to break down due to the high-strength vortices generated by the plate's edges. The initial roll-up is extremely similar between both visualization techniques, and the shape of the vortex ring is extremely well captured by the simulations despite the minimal domain size.

\section{Developed vortex-wake applications}

Next, we study the accuracy of the new Biot-Savart conditions on flow cases with vortex wakes that are truncated by the domain boundary, introducing an additional modeling error. We use two canonical developed vortex-wake examples: (1) the flow around a sphere at $\text{Re}=3700$, and (2) the reversed \emph{van K\`arm\`an} (or propulsive) wake generated by a heaving 2D airfoil over a range of Strouhal numbers.

\subsection{3D flow around a sphere}

\begin{figure}
    \centering
    \begin{subfigure}{.8\textwidth}
        \centering
        \includegraphics[width=\textwidth]{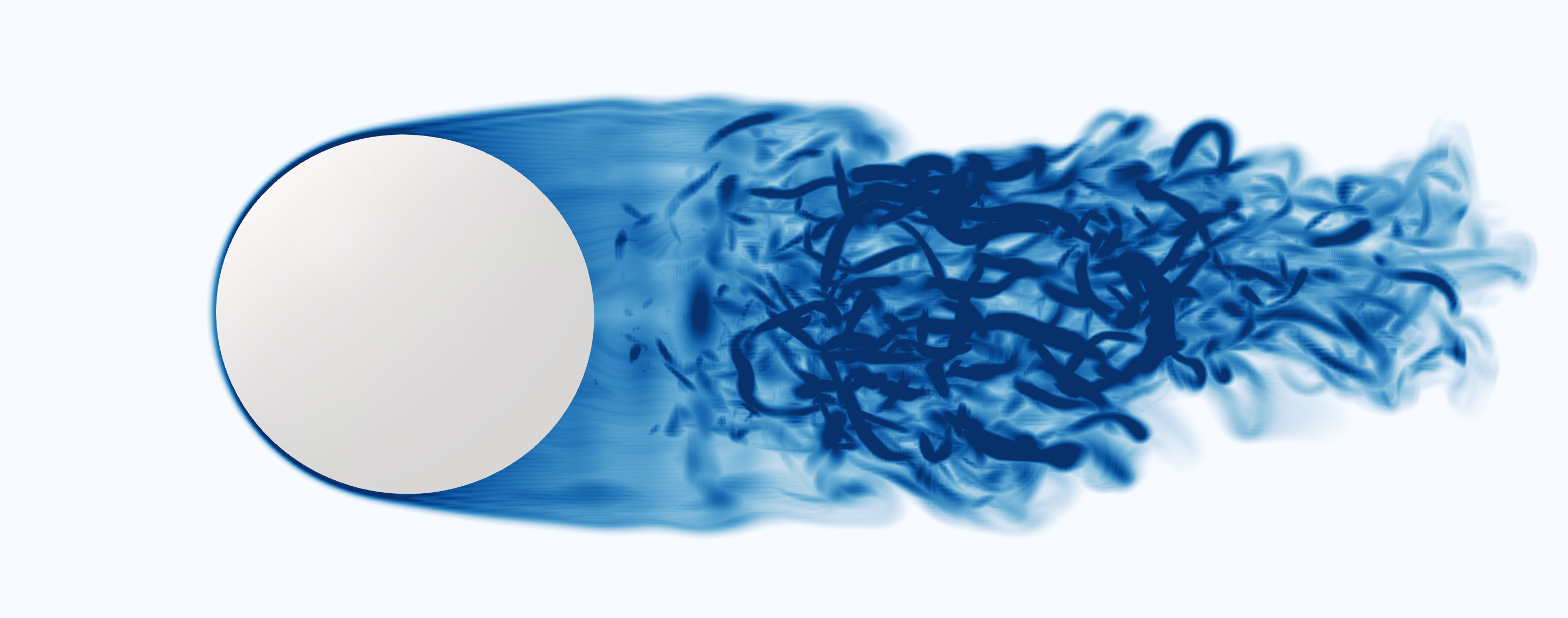}
    \end{subfigure}%
    \\
    \begin{subfigure}{.4\textwidth}
        \centering
        \includegraphics[width=\textwidth]{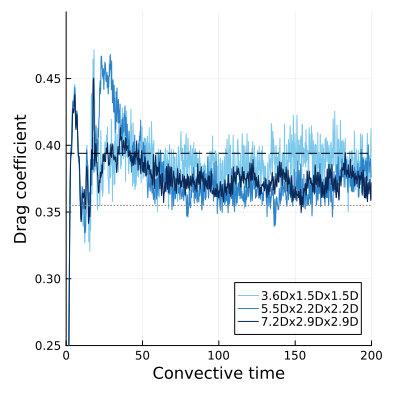}
    \end{subfigure}%
    \hspace{6mm}
    \begin{subfigure}{.4\textwidth}
        \centering
        \includegraphics[width=\textwidth]{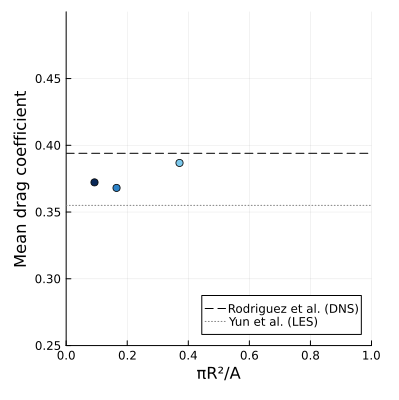}
    \end{subfigure}
    \caption{Sphere flow vortex wake and drag coefficient at $Re = UR/\nu= 3700$ with a resolution of 44 cells per radius. (top) Wake visualized by a volumetric rendering of $|\omega|R/U$ at time $tU/R=53$ with the full $3.6D\times1.5D\times1.5D$ domain shown. Instantaneous (right) and time-averaged (left) drag coefficients for three domains, where $A$ is the domain frontal area. The time-averages drag is obtained by averaging the last 100 convective times.}
    \label{fig:sphere}
\end{figure}

The flow around a sphere is used to validate the coupled momentum-Biot-Savart update for developed 3D flows. A sphere of radius $R$ is immersed in a viscous incompressible flow. The Reynolds number is set to $Re=UR/\nu=3700$, we use a resolution of 44 cells per radius and use a domain size of $5C/2 \times C \times C$ cells varied over $C\in[128,192,256]$. The flow is impulsively started from rest to a uniform velocity $U$. 

We perform the simulations until convective time $tU/R=200$ and Figure~\ref{fig:sphere} shows the drag coefficient when using the Biot-Savart boundary condition is insensitive to the domain size - even for domains only 50\% wider than the sphere diameter. The averaged drag over the last 100 convective time for all blockage ratios compares favorably to the DNS and LES simulation results obtained in \cite{Yun2006VorticalNumbers, Rodriguez2011Direct3700}. 

\subsection{Wake behind a heaving foil}

The previous results suggest that the wake history on the immersed body is very small for drag wakes. We next continue our analysis with thrust wakes, where the influence of wake history is expected to be stronger.

We immerse a rigid foil of length $L$ moving with velocity $U$ in a viscous fluid. In addition to its steady forward speed, the foil undergoes a pure heave motion of amplitude $h_0$
\begin{equation}
    y(t) = h_0 \sin(2\pi f t)
\end{equation}
where $h_0/L=0.5$ is the non-dimensional amplitude, and the frequency is $f = \text{St}\,U/h_0$. For Strouhal number $\text{St}<0.5$, the airfoil generates thrust, and almost no net lift, and for $\text{St}>0.6$, the wake generated by this flapping foil is strongly asymmetric and generates a significant net lift force. We perform simulations of this system for a range of Strouhal numbers and compare the mean lift force and wake obtained on a large domain ($30L\times20L$) with standard reflective boundary conditions to the ones obtained on two different domains of size ($6L\times4L$) and ($12L\times8L$) that use the Biot-Savart boundary conditions. We simulate the flow around this heaving airfoil for 100 convective times to ensure that the wake stabilizes, and we gather pressure and viscous forces for the last 20 convective times. Mean lift forces are obtained by time-averaging these forces over the last 20 convective times.

\begin{figure}
    \centering
    \begin{subfigure}{.48\textwidth}
        \centering
        \includegraphics[trim={2.8cm 2cm 4cm 2cm},clip,width=\textwidth]{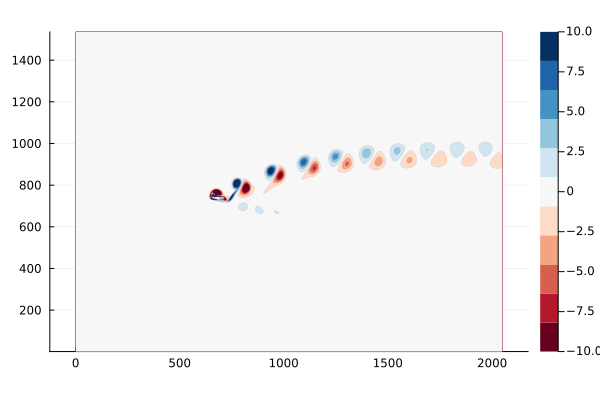}
        \caption{Reflection ($30L\times20L$)}
    \end{subfigure}%
    \hspace{0.1cm}
    \begin{subfigure}{.48\textwidth}
        \centering
        \includegraphics[trim={2.8cm 2cm 4cm 2cm},clip,width=\textwidth]{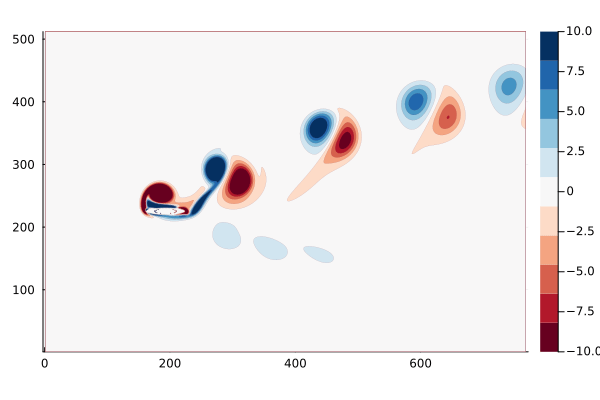}
        \caption{Reflection ($30L\times20L$) - near wake}
    \end{subfigure}
    \begin{subfigure}{.48\textwidth}
        \centering
        \includegraphics[trim={2.8cm 2cm 4cm 2cm},clip,width=\textwidth]{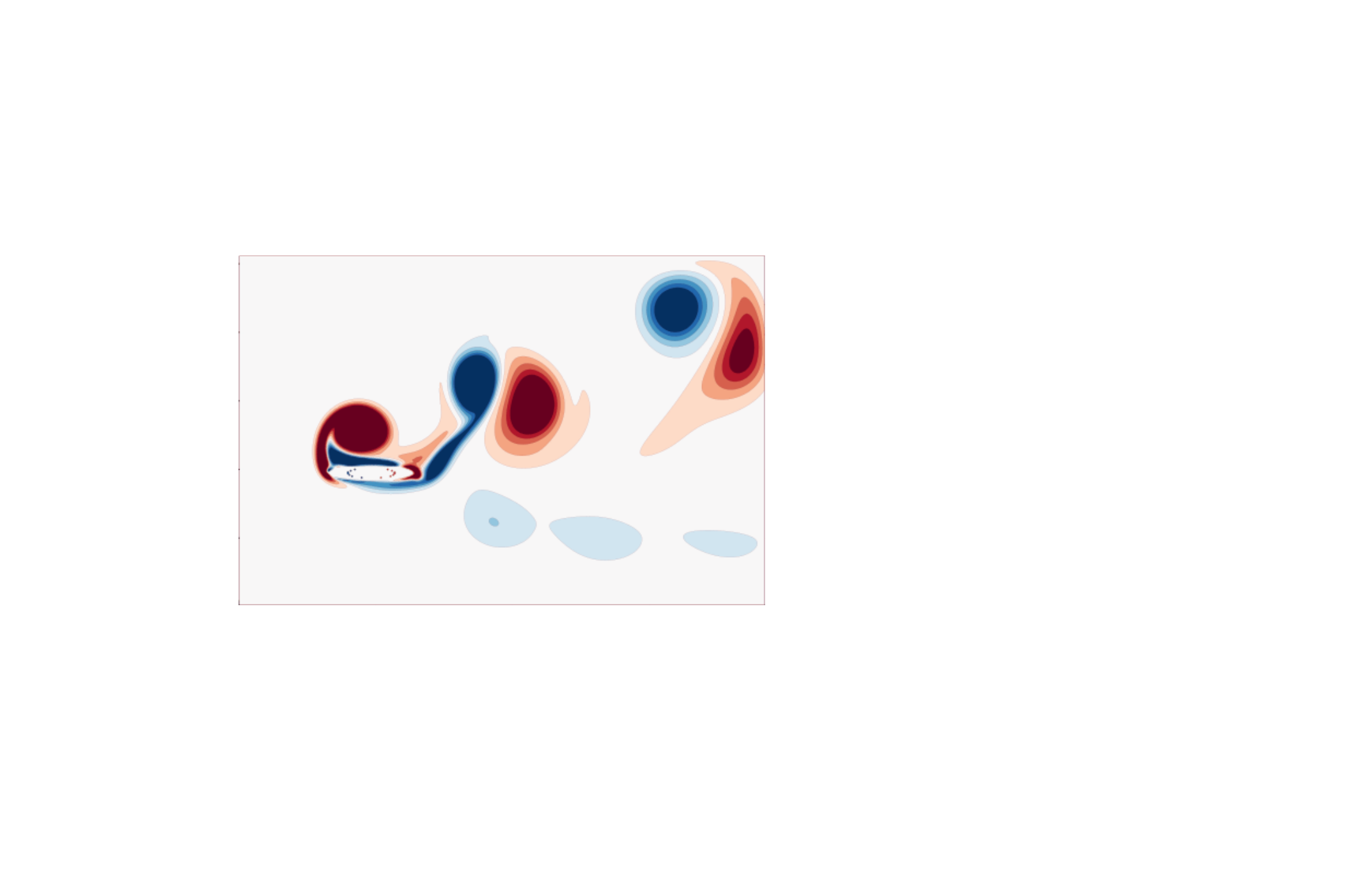}
        \caption{Biot-Savart ($6L\times4L$)}
    \end{subfigure}%
    \hspace{0.1cm}
    \begin{subfigure}{.48\textwidth}
        \centering
        \includegraphics[trim={2.8cm 2cm 4cm 2cm},clip,width=\textwidth]{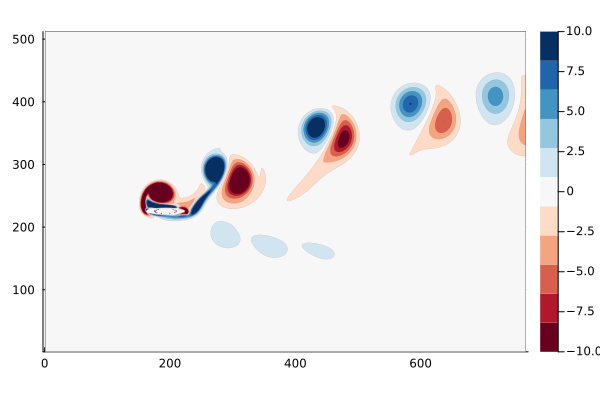}
        \caption{Biot-Savart ($12L\times8L$)}
    \end{subfigure}
    \caption{Snapshot of the deflected wake behind an airfoil at $\text{St}=0.6$ and $\text{Re}=100$ for an amplitude to chord ratio $h_0/L=1.0$. The upper left panel $(a)$ shows the far wake obtained with the reflective boundary condition in a large domain, the upper right panel $(b)$ shows a zoom onto the airfoil and the near wake, the bottom left $(c)$ panel shows the entire (small) Biot-Savart domain used for the computation and the lower right panel $(d)$ shows the entire (larger) Biot-Savart domain. Vorticity is shown as 10 equally spaced isocontours in the interval $\omega L/U \pm 10$.}
    \label{fig:deflected_wake}
\end{figure}

\begin{figure}
    \centering
    \begin{subfigure}{.5\textwidth}
        \centering
        \includegraphics[trim={0 0 0 0},clip,width=\textwidth]{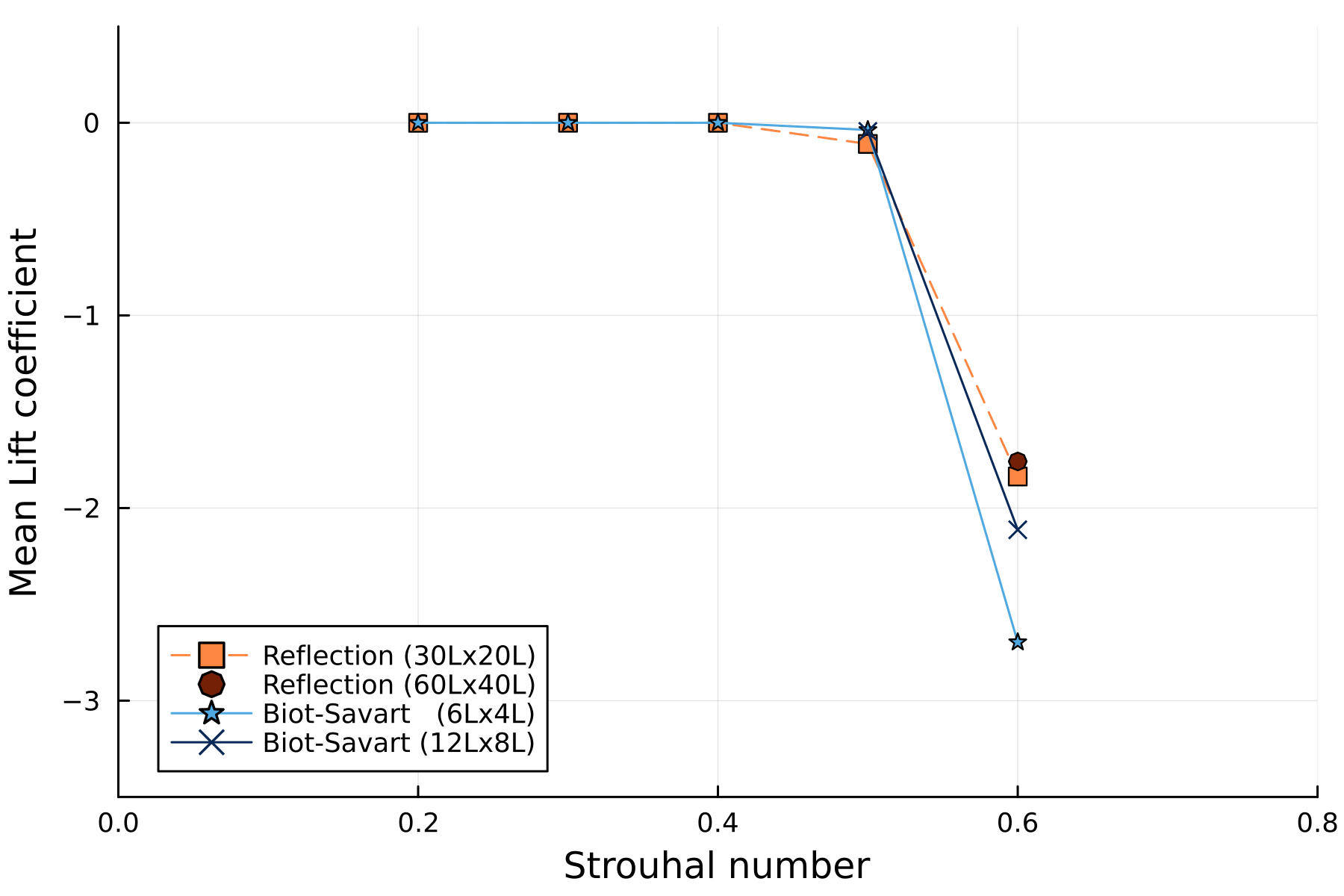}
    \end{subfigure}%
    \begin{subfigure}{.5\textwidth}
        \centering
        \includegraphics[trim={0 0 0 0},clip,width=\textwidth]{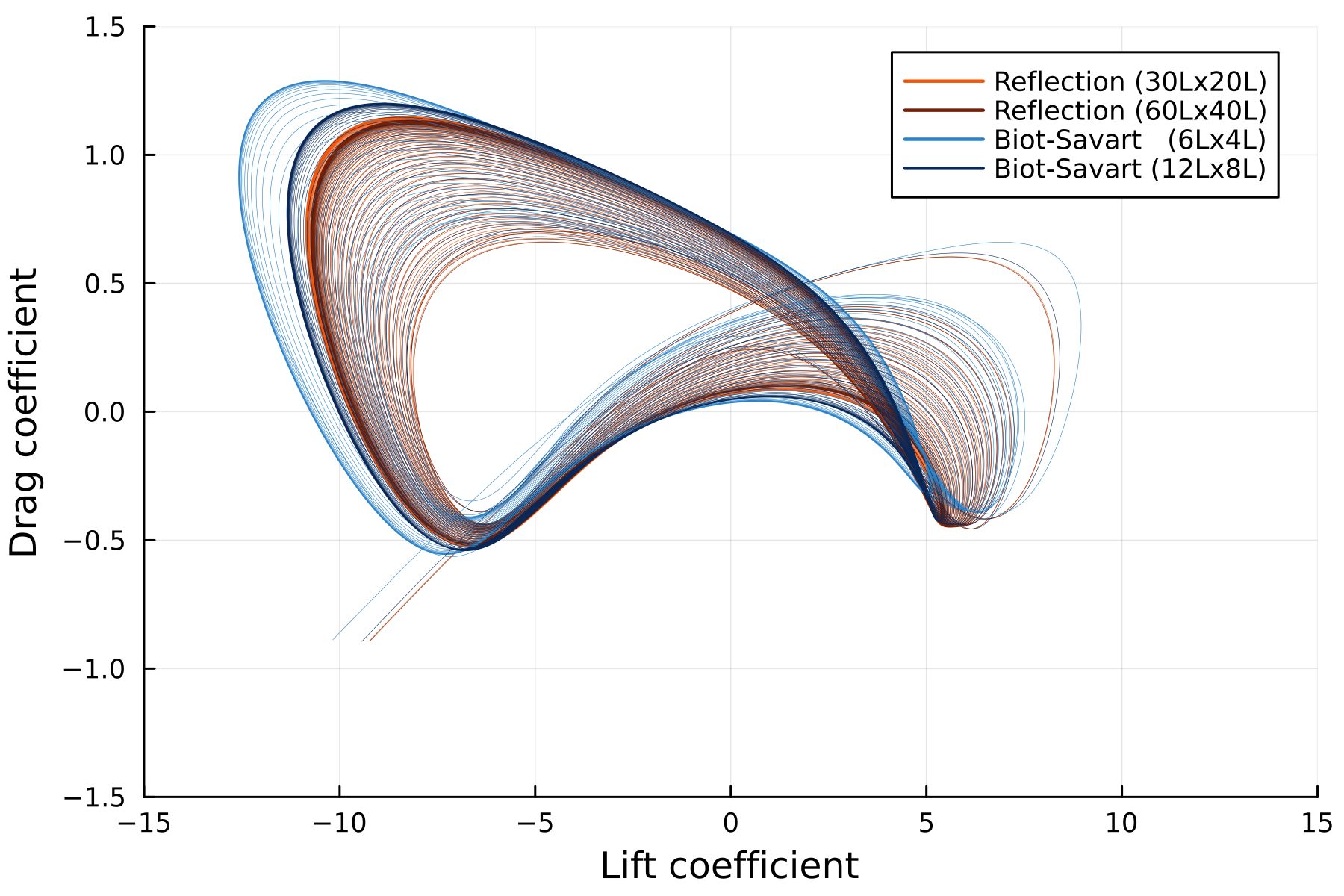}
    \end{subfigure}
    \caption{(left) Mean lift force acting on the airfoil at $\text{Re}=100$, for a range of Strouhal numbers for the Biot-Savart and the reflective boundary condition  and (right) \emph{Poincar\'e} map of the lift and drag coefficients for $\text{St}=0.6$}
    \label{fig:deflected_wake_2}
\end{figure}

We observe almost no difference between the domain sizes for $\text{St}<0.5$. The wake is symmetric, and the airfoil produces zero mean lift forces; Fig.~\ref{fig:deflected_wake_2}(a). For $\text{St}\ge 0.5$, we observe wake deflection due to the high frequency and large circulation vortices pairing up into vortex-couples as seen in Fig.~\ref{fig:deflected_wake} for $\text{St}=0.6$. Fig.~\ref{fig:deflected_wake_2}(b) shows a \emph{Poincar\'e map} for the lift and drag coefficient of the heaving airfoil at $\text{St}=0.6$, with the asymmetry in the lift resulting in a net mean lift coefficient around $C_L=-1$ for the very large domain with reflection BCs. The new Biot-Savart boundary conditions predict wake deflection for $\text{St}\ge 0.5$ correctly, but truncating this sensitive vortex wake causes the space between the vortex couples to reduce, Fig.~\ref{fig:deflected_wake}(c,d). Using a domain extending only four chord lengths behind the foil results in a 10-40\% over-prediction of the measured forces, while doubling the domain recovers the correct spacing and forces to within 5\%.

We expect that in any case where wake history effects are highly dominant, the new Biot-Savart boundary condition will require a domain large enough for this wake to develop stable dynamics before exiting the domain. However, we should note that the case of 2D deflected thrust wakes is somewhat pathological, depending on long-range vortex interactions only possible in 2D flows. Indeed, experiments on finite foils have shown that these high-lift deflected jets cannot form for finite wings, see \cite{Calderon2014OnWings, Godoy-Diana2008TransitionFoil}.
 
\section{Conclusion}

In this paper, we present a novel coupling to impose unbounded boundary conditions on viscous incompressible flow simulations. The method relies on a Biot-Savart integral of the vorticity inside the domain to set the velocity boundary conditions on the domain boundaries. We couple these boundary conditions with an explicit projection scheme to solve the momentum and continuity equations in their primitive variable form. Critically, we demonstrate that when we apply the Biot-Savart equation to the domain boundaries, a coupling between every point on immersed solid boundaries and the domain boundary is introduced. This coupling results in a dense pressure equation that we solve using iterationss on the projection and Biot-Savart update steps, allowing a standard Poisson solver to be used. This approach also makes the solution extremely robust, enabling simple and high-speed FMM variants to be used without sacrificing solution accuracy. In this work, we use a multi-level clustering method for the $N$ Eulerian-grid vortex sources, and use this to develop a simple Fast Multi-level Method (FM$\ell$M) for the velocity reconstruction with $O(N)$ complexity that can be run in parallel on CPU and GPU architectures. We show that the error introduced by this multilevel approach is bounded and that its influence on the resulting flow field is minimal because of the iterative partitioned approach used for the pressure.

With different examples of flow with confined vorticity, we show that the novel boundary conditions allow for a significant reduction in the size of the computational domain without any loss in accuracy. The method exactly captures added-mass force on an accelerated disk with a computational domain only extending $1/2$ diameter away from the plate. This domain is also large enough to perfectly replicate the theoretical and previous numerical studies on an impulsively started circular cylinder. In comparison, a domain more than 100 times larger is required when using reflection boundary conditions.

Surprisingly, the method demonstrates its robustness for flows with developed vortex wakes. The developed flow on a sphere is found to be largely independent of the domain size, even for blockage ratios greater than 40\%. 
Lastly, we show that the method is limited when a strong wake history influences the flow in the near wake. For a high-amplitude, high-Strouhal number ($St\ge0.5$) heaving airfoil, we show that capturing the deflected wake of the flow requires a significantly larger domain, implying that the computational advantage of the new boundary condition is reduced.

The generality and accuracy of the new approach, even when simulating flows with vortex wakes truncated by the domain boundaries, opens a large range of potential applications. In addition to the welcomed speed-up enabled by using minimal computational domains, the continued accuracy of the simulations challenges the notion of the importance of far-field wake dynamics on the near-body flow, providing compelling evidence to the contrary.


\appendix

\section{Multilevel error analysis}\label{A0}

\subsection{Analytic error bound}

We derive the error induced by this multilevel approach, starting with the 2D-case and extending the result to 3D. The velocity at location $\vec x$ induced by the circulation from a single cell on level $i+1$ is given by (see Fig.~\ref{Fig_2})
\begin{equation}\label{eq:induced_1}
    \vec{u}^{(i+1)}(\vec{x}) = \frac{\vec{r}}{2\pi|\vec{r}|^2}\times\vec{e}_z\mathcal{P}^{(i)\to(i+1)}\Gamma^{(i)}
\end{equation}
where $\vec{r}=\vec{y}-\vec{x}$, $y$ is the center of the cell, and $\vec{e}_z$ is the out-of-plane unit vector. The velocity induced by the four corresponding cells on level $i$ is
\begin{equation}
    \vec{u}^{(i)}(\vec{x}) = \mathcal{P}^{(i)\to(i+1)}
    \left(\frac{\vec{r}_j}{2\pi|\vec{r}_j|^2}\times\vec{e}_z\Gamma^{(i)}_j\right),
\end{equation}
where $\vec{r}_j=\vec{y}_j-\vec{x}$ and $y_j$ are the four cell centers. We can further decompose this expression by noting that $\vec{r}_j=\vec{y}-\vec{x}-\vec{\delta r}_j=\vec{r}-\vec{\delta r}_j$.
For sufficiently large $\vec{r}$ we can assume that $|\vec{r}|\approx|\vec{r}-\vec{\delta r}_j|$ 
which allows us to separate the kernel's influence into two parts
\begin{equation}
    \vec{u}^{(i)}(\vec{x}) = \mathcal{P}^{(i)\to(i+1)}    \left(\frac{\vec{r}}{2\pi|\vec{r}|^2}\times\vec{e}_z\Gamma^{(i)}_j-\frac{\vec{\delta r}_j}{2\pi|\vec{r}|^2}\times\vec{e}_z\Gamma^{(i)}_j\right).
\end{equation}
After rearrangement, the first term equals Eq.~\ref{eq:induced_1}. The second term is the error due to the multilevel approximation. Its magnitude is
\begin{equation}
    \varepsilon(\vec{x})^{(i)} = \mathcal{P}^{(i)\to(i+1)}\frac{\vert\vec{\delta r}_j\vert}{2\pi|\vec{r}|^2}\Gamma^{(i)}_j,
\end{equation}
as $\vec{e}_z$ is a unit vector. On uniform Cartesian meshes, $\vert\vec{\delta r}_j^{(i)}\vert\sim2^{l}$, while $|\vec{r}|^2 \sim (2^l\tilde{S})^2$. An upper bound for the total error in the induced velocity at a point $\vec x$ due to the multilevel evaluation of the Biot-Savart integral is therefore
\begin{equation}
    \varepsilon(\vec{x}) \approx \frac{1}{2\pi\tilde{S}^2}\sum_{i=2}^{l}\frac{1}{2^i}\mathcal{P}^{(i)\to(i+1)}
    \Gamma^{(i)}.
\end{equation}
which decays as $\tilde{S}^{-2}$. The content of each sum is a decreasing fraction of the total vorticity contained in the domain $\mathcal{D}^{(i)}$. As expected, taking $\tilde{S}>\text{Domain}$ yields $\varepsilon(\vec{x}) = 0$, and as $\Gamma^{(i)} \to 0, \varepsilon(\vec{x})\to 0$, finally as $\tilde{S}\to0, \varepsilon(\vec{x}) \to \infty$. Similarly, in 3D the upper bound for the error is given by
\begin{equation}
    \varepsilon(\vec{x}) \approx \frac{1}{4\pi\tilde{S}^3}\sum_{i=2}^{l}\frac{1}{2^{2i}}\mathcal{P}^{(i)\to(i+1)}
    \Gamma^{(i)}.
\end{equation}
which decays as $\tilde{S}^{-3}$. Similarly, the content of the sum is a (even smaller) decreasing fraction of total vorticity contained in each level, improving the method's accuracy in 3D compared to 2D.

\subsection{2D vortex reconstruction example}

We next use an analytic 2D vortex field to demonstrate the error in the reconstruction method. The velocity field generated by the Lamb-Chaplygin dipole is defined by the scalar stream function
\begin{equation}
    \psi(r) = \begin{cases}
    \frac{-2UJ_1(\beta r)}{\beta J_0(\beta R)}, \quad \text{for} \quad r < R,\\
    U\left(\frac{R^2}{r}-r\right), \quad \text{for} \quad r \ge R,    \end{cases}
\end{equation}
where $J_0$ and $J_1$ are the zeroth and first Bessel functions of the first kind, respectively, and $\beta$ is a parameter with the value $\beta=1.2197\pi/R$. The 2D velocity field is given by $\vec{u} = -\vec{e}_z\times \nabla \psi$, and the core of the vortex is defined by the isoline $\psi(r=R)=0$.

\begin{figure}
    \centering
    \begin{subfigure}{.5\textwidth}
        \centering
        \includegraphics[width=\textwidth]{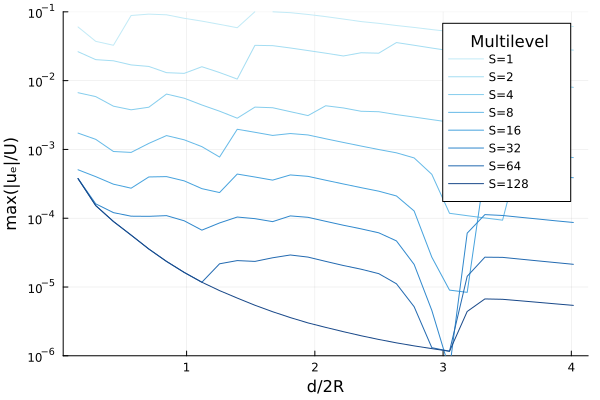}
    \end{subfigure}%
    \begin{subfigure}{.5\textwidth}
        \centering
        \includegraphics[width=\textwidth]{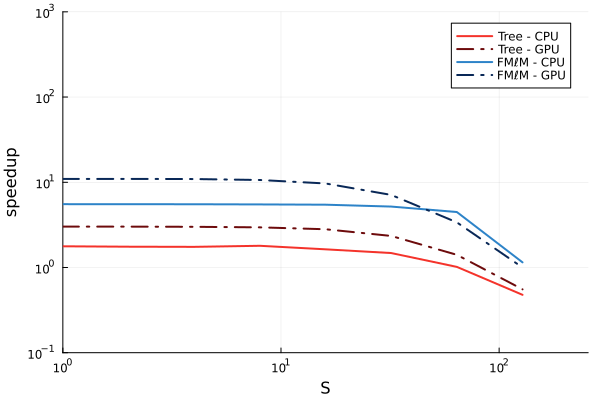}
    \end{subfigure}
    \caption{Maximum error in the reconstructed velocity field generated by a 2D \emph{Lamb-Chaplygin} vortex dipole for various kernel sizes and distances from the vortex core (left). Corresponding speed-up obtained by reducing the kernel size for the multilevel algorithm (right). The mesh has dimensions $[2^8,2^8]$.}
    \label{fig:error_lamb_2}
\end{figure}

We reconstruct the velocity field outside the vortex core using the internal vorticity and our multilevel Biot-Savart method and compare that to the analytic result. We reconstruct the velocity on a domain of size $[2^8,2^8]$. The dependency of the absolute error in the reconstructed velocity field on the half-width $\tilde S$ of the subdomains are shown in Fig.~\ref{fig:error_lamb_2}(a). Setting $\tilde S$ to the domain size $\cal D$ eliminates the source clustering error, but the result is not exact since the analytic vorticity distribution is sampled on a finite grid of size $\Delta x=0.05
R$. This error depends primarily on the distance from the vortex, Fig.~\ref{fig:error_lamb_2}(a). Progressively reducing the subdomain size introduces the multilevel error quantified in appendix A, which is only weakly dependent on the distance to the vortex. Fig.~\ref{fig:error_lamb_2}(b) reports the measured computational speed-up achieved by the multilevel method relative to setting $\tilde S=\Omega$, demonstrating up to 300$\times$ acceleration.

\subsection{3D vortex reconstruction example}

\begin{figure}
    \centering
    \begin{subfigure}{.5\textwidth}
        \centering
        \includegraphics[width=\textwidth]{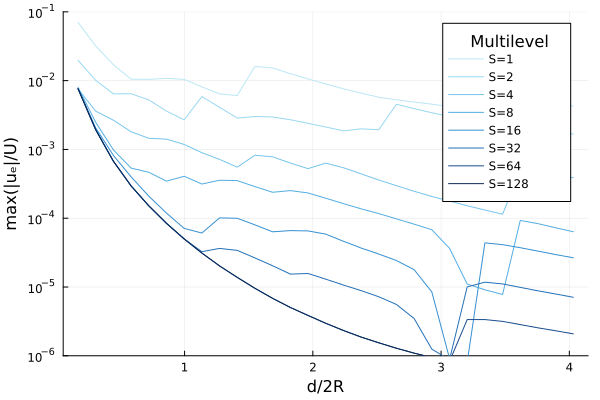}
    \end{subfigure}%
    \begin{subfigure}{.5\textwidth}
        \centering
        \includegraphics[width=\textwidth]{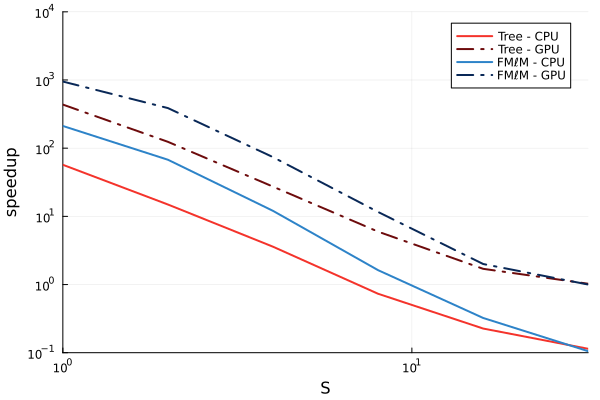}
    \end{subfigure}
    \caption{Maximum error in the reconstructed velocity field generated by a 3D \emph{Hill} vortex for various kernel sizes and distances from the vortex core (left). Corresponding speed-up obtained by reducing the kernel size for the multilevel algorithm (right). The mesh has dimensions $[2^8,2^8,2^8]$.}
    \label{fig:error_hill_3}
\end{figure}

This validation is repeated in 3D using the classical \emph{Hill vortex}, whose radial and tangential velocity components are given by
\begin{align}
    v_r &= \frac{1}{r^2\sin\theta}\frac{\partial\psi}{\partial\theta}, \qquad v_\theta = -\frac{1}{r\sin\theta}\frac{\partial\psi}{\partial r}.
\end{align}
and the stream function 
\begin{equation}
    \psi(r,\theta) = \begin{cases}
    -\frac{3U}{4}\left(1-\frac{r^2}{R^2}\right)r^2sin^2\theta \quad \text{for} \, r \le R,\\
     \,\,\frac{U}{2}\left(1-\frac{r^3}{R^3}\right)r^2sin^2\theta \quad \text{for} \, r > R.
    \end{cases}
\end{equation}
with the corresponding errors and speed-up shown in Figure~\ref{fig:error_hill_3}. Note that the analytic vorticity of the hill vortex is discontinuous, increasing the truncation error near the vortex compared to the smooth 2D Lamb dipole.

The multilevel error decreases extremely quickly as $\tilde S$ increases for three-dimensional flows. This is due to the scaling of the Biot-Savart kernel itself. Similarly, the speed-up is much faster in 3D because each doubling gathers 8 cells instead of 4, with a maximum observed speed-up of nearly 20,000$\times$ using the multilevel method.

\section{Accounting for symmetries in the Biot-Savart integral}
\label{A1}

\begin{figure}
    \centering
    \def\svgwidth{0.5\columnwidth}
    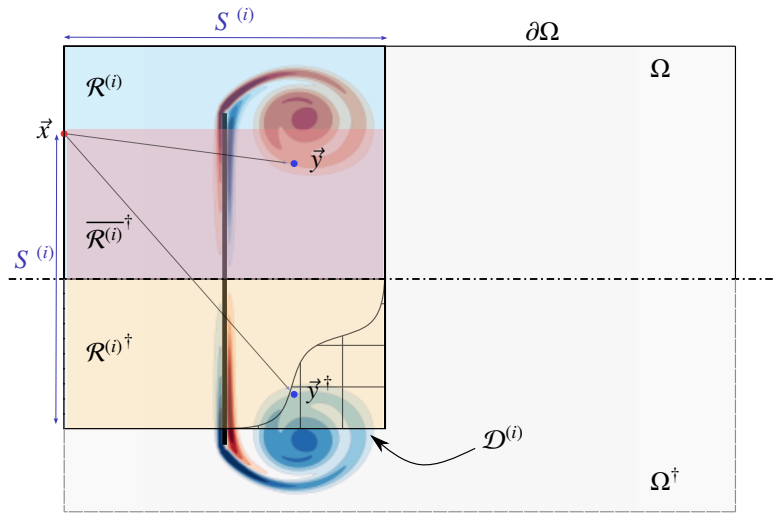
    \caption{Schematic of the approach used to account for symmetries in the multilevel evaluation of the Biot-Savart integral. The induced velocity at a point $\vec{x}$ includes the contribution from $\vec{y}$ and its image point $\vec{y}^\dagger$. The computational domain $\Omega$ and its image $\Omega^\dagger$ intersect with a multilevel domain $\mathcal{D}^{(i)}$ of half-width $S^{(i)}$ producing the regions $\mathcal{R}^{(i)}$, ${\mathcal{R}^{(i)}}^\dagger$ and $\overline{\mathcal{R}^{(i)}}^\dagger$.}
    \label{a1:fig_1}
\end{figure}

Symmetries can be accounted for in the multilevel evaluation of the Biot-Savart integral Eq.~\ref{eq:Biot}. We focus on a description of the 2D case; the extension to the 3D is trivial. For a flow with symmetries, like the 2D flow around a flat plate, see Figure~\ref{a1:fig_1}, whose axis of symmetries is represented by the dash-dotted line, the computational domain $\Omega$ only models the upper part of the plate. The domain $\Omega^\dagger$ is the image of the computational domain $\Omega$ about the symmetry axis.

Evaluation of the induced velocity at a point on the boundary of the computational domain $\vec{x}\in\partial\Omega$ (apart from points on the symmetry plane which have the reflection condition applied to them) on the $i$th-level is given by
\begin{equation}\label{a1;eq}
     {\vec u}({\vec x}) = \int_{\mathcal{R}^{(i)}} K_{n}({\vec x} - \vec{y})\times \hat{e}_z\omega({\vec y})\text{ d}\vec{y} + \int_{{\mathcal{R}^{(i)}}^\dagger} K_{n}({\vec x} - \vec{y}^\dagger)\times \hat{e}_z\omega^\dagger({\vec y}^\dagger)\text{ d}\vec{y}^\dagger
\end{equation}
where $\mathcal{R}^{(i)}\equiv\mathcal{D}^{(i)}\cap\Omega$ and ${\mathcal{R}^{(i)}}^\dagger\equiv\mathcal{D}^{(i)}\cap\Omega^\dagger$. Additionally, $\overline{\mathcal{R}^{(i)}}^\dagger\subset\mathcal{R}^{(i)}$ is the images of ${\mathcal{R}^{(i)}}^\dagger$ in $\Omega$ and ${\vec{y}}^\dagger$ is the images of $\vec{y}$ in $\Omega^\dagger$, respectively.

Points inside ${\mathcal{R}^{(i)}}$ (blue region) are evaluated on the computational domain $\Omega$. For points in ${\mathcal{R}^{(i)}}^\dagger$ (orange region), the image vorticity is computed by applying the symmetric condition to the vorticity in $\overline{\mathcal{R}^{(i)}}^\dagger$ (pink region)
\begin{equation}
    \omega^\dagger(\vec{y}^\dagger) = -\omega(\vec{y}), \quad\quad \vec{y}\in\overline{\mathcal{R}^{(i)}}^\dagger,
\end{equation}
for 2D flows; for 3D flows, computing the image vorticity vector requires computing its reflection about the symmetry plane. Substituting in Eq.~\ref{a1;eq} yields an integral only over the region $\mathcal{R}^{(i)}$
\begin{equation}
     {\vec u}({\vec x}) = \int_{\mathcal{R}^{(i)}} \left(K_{n}({\vec x} - \vec{y})-K_{n}({\vec x} - \vec{y}^\dagger)\right)\times \hat{e}_z\omega({\vec y})\text{ d}\vec{y},
\end{equation}
where the second term in the parenthesis only appears if $\mathcal{D}^{(i)}$ overlaps with the domain's image $\Omega^\dagger$. A similar expression is found for 3D flows. As Eq.~\theequation~ solely relies on the vorticity in the computational domain $\Omega$, no additional computations are required, and accounting for the image's induced velocity simply requires accounting for its influence through the kernel $K_n$.

 \bibliographystyle{elsarticle-num} 
 \bibliography{references}





\end{document}